\documentclass[fleqn,usenatbib]{mnras}

\usepackage{newtxtext,newtxmath}
\usepackage{ulem}

\usepackage[T1]{fontenc}

\DeclareRobustCommand{\VAN}[3]{#2}
\let\VANthebibliography\thebibliography
\def\thebibliography{\DeclareRobustCommand{\VAN}[3]{##3}\VANthebibliography}

\usepackage{graphicx}	
\usepackage{amsmath}	

\title[Gaia21bty - a new eruptive star]{Gaia21bty: An EXor lightcurve exhibiting an FUor spectrum}

\author[M. Siwak et al.]{Micha{\l} Siwak$^{1,2}$\thanks{E-mail: michal.siwak@csfk.org},
Lynne A. Hillenbrand$^{3}$,
\'{A}gnes K\'{o}sp\'{a}l$^{1,2,4,5}$,
P\'{e}ter \'{A}brah\'{a}m$^{1,2,4}$,
Teresa Giannini$^{6}$,\newauthor
Kishalay De$^{7}$,
Attila Mo\'or$^{1,2}$,
M\'{a}t\'{e} Szil\'{a}gyi$^{1,2,4}$,
Jan Jan\'{i}k$^{8}$,
Chris Koen$^{9}$,
Sunkyung Park$^{1,2}$,
Zs\'ofia Nagy$^{1,2}$,\newauthor
Fernando Cruz-Sáenz de Miera$^{1,2}$,
Eleonora Fiorellino$^{10,1,2}$,
G\'{a}bor Marton$^{1,2}$,
M\'aria Kun$^{1,2}$,
Philip W. Lucas$^{11}$,\newauthor
Andrzej Udalski$^{12}$,
and Zs\'ofia Marianna Szab\'o$^{13,14,1,2}$
\\
$^{1}$Konkoly Observatory, Research Centre for Astronomy and Earth Sciences, E{\"o}tv{\"o}s Lor\'and Research Network (ELKH), Hungarian Academy of Sciences,\\~~~Konkoly-Thege Mikl\'os \'ut 15--17, 1121 Budapest, Hungary\\
$^{2}$CSFK, MTA Centre of Excellence, Budapest, Konkoly Thege Mikl\'os \'ut 15-17, H-1121, Hungary\\
$^{3}$Department of Astronomy, California Institute of Technology, Pasadena CA 91125, USA\\
$^{4}$ELTE E\"otv\"os Lor\'and University, Institute of Physics, P\'azm\'any P\'eter s\'et\'any 1/A, 1117 Budapest, Hungary\\
$^{5}$Max Planck Institute for Astronomy, Königstuhl 17, 69117 Heidelberg, Germany\\
$^{6}$INAF-Osservatorio Astronomico di Roma, via di Frascati 33, 00040, Monte Porzio Catone, Italy\\
$^{7}$Kavli Institute for Astrophysics and Space Research, Massachusetts Institute of Technology, Cambridge, MA 02139, USA\\
$^{8}$Department of Theoretical Physics and Astrophysics, Masaryk University, Kotl\'{a}\v{r}sk\'{a} 2, Brno, CZ-611 37, Czech Republic\\
$^{9}$Department of Statistics, University of the Western Cape, Private Bag X17, Bellville, 7535 Cape, South Africa\\
$^{10}$INAF-Osservatorio Astronomico di Capodimonte, via Moiariello 16, 80131 Napoli, Italy\\
$^{11}$Centre for Astrophysics, University of Hertfordshire, College Lane, Hatfield, AL10 9AB, UK\\
$^{12}$Astronomical Observatory, University of Warsaw, Al. Ujazdowskie 4, 00-478 Warszawa, Poland\\
$^{13}$Max-Planck-Institut für Radioastronomie, Auf dem Hügel 69, 53121 Bonn, Germany \\
$^{14}$Scottish Universities Physics Alliance (SUPA), School of Physics and Astronomy, University of St Andrews, North Haugh, St Andrews, KY16 9SS, UK
}

\date{Accepted 2023. Received 2023; in original form 2023}

\pubyear{2023}

\begin{document}
\label{firstpage}
\pagerange{\pageref{firstpage}--\pageref{lastpage}}
\maketitle

\begin{abstract}
Gaia21bty, a pre-main sequence star that previously had shown aperiodic dips in its light curve, underwent a considerable $\Delta G\approx2.9$~mag brightening that occurred over a few months between 2020 October -- 2021 February. 
The Gaia lightcurve shows that the star remained near maximum brightness for about $4-6$ months, and then started slowly fading over the next 2 years, with at least three superimposed $\sim$1~mag sudden rebrightening events. Whereas the amplitude and duration of the maximum is typical for EXors, optical and near-infrared spectra obtained at the maximum are dominated by features which are typical for FUors. Modelling of the accretion disc at the maximum indicates that the disc bolometric luminosity is 43~L$_{\odot}$ and the mass accretion rate is $2.5\times10^{-5}$~M$_{\odot}$~yr$^{-1}$, which are typical values for FUors even considering the large uncertainty in the distance ($1.7_{-0.4}^{+0.8}$~kpc). Further monitoring is necessary to understand the cause of the quick brightness decline, the rebrightening, and the other post-outburst light changes, as our multi-colour photometric data suggest that they could be caused by a long and discontinuous obscuration event. We speculate  that the outburst might have induced large-scale inhomogeneous dust condensations in the line of sight leading to such phenomena, whilst the FUor outburst continues behind the opaque screen.
\end{abstract}




\section{Introduction}

FU~Ori-type stars (FUors) were recognized by \citet{Herbig1977_ApJ217693H} as classical T Tauri-type stars (CTTS) undergoing enhanced disc-matter accretion for a few decades or longer. In addition to FUors, \citet{Herbig1989} established the group of EX~Lupi-type stars (EXors), showing less dramatic eruptive events lasting considerably shorter, but re-appearing on the time scale of a few -- a dozen of years. This bimodal classification was recently blurred upon the discovery of eruptive stars showing the properties of both FUors and EXors \citep{Audard_2014prpl.conf387A}, e.g. V1647~Ori \citep{Andrews2004, Acosta2007, Muzerolle2015} and V899~Mon \citep{Ninan2015, Park2021}.
A recent summary of the enhanced accretion process across the whole range of pre-main-sequence (PMS) evolutionary stages and an attempt of unification of the zoo of accompanying photometric and spectroscopic features were presented by \citet{Fischer2022}, and in a modest form by \citet{Contreras-Pena2023}.

Studies show also that eruptive events are much more common for Class~I than Class~II objects \citep{Contreras-Pena2019, Guo2021, Fiorellino2023}. According to e.g. \citet{Contreras-Pena2019}, all PMS stars should undergo at least a dozen of enhanced accretion events. This provides evidence for episodic accretion as the preferred model to explain the observed protostellar luminosity spread \citep{Fischer2022}. As enhanced accretion has a major impact on the inner disc temperature causing the expansion of the snow line of various volatiles \citep{Cieza2016}, multiwavelength studies of outbursting young stars are also crucial for understanding the chemical evolution of protoplanetary discs and eventually also the composition and morphology of planetary systems, including our Solar System \citep{Abraham2009_Natur459224A, Abraham2019_ApJ887156A, Hubbard2017, Molyarova2018, Wiebe2019}.

The recently most recognised list of about 40 young eruptive stars compiled by \citet{Audard_2014prpl.conf387A} expands constantly.
The ground-based All-Sky Automated Survey for Supernovae (ASAS-SN, \citealt{asas-sn-1,asas-sn-2}) and the Zwicky Transient Facility (ZTF, \citealt{masci2018,Masci2019_PASP131a8003M}) have discovered several outbursting stars. These include the EXor/FUor-type object ASASSN-13db \citep{Holoien2014, Sicilia2017}, PTF14jg tentatively identified as a FUor-type object \citep{Hillenbrand2019b}, and so far unclassified strong but relatively short outburst in ASASSN-15qi \citep{Herczeg2016}. 
Discovery of 106 eruptive stars was reported in \cite{Contreras-Pena2017a,Contreras-Pena2017b} during the VISTA Variables in the Via Lactea (VVV, \citealt{Minniti2010}) survey conducted in near-infrared bands.

The multi-purpose Gaia spacecraft observes each part of the sky in broad-band filters several dozen times a year \citep{Gaia2016}. After a few years of continuous observations, Gaia became an effective tool for catching long-term brightness changes in various astronomical objects down to 21~mag. The community is notified about significant brightness variations via the Gaia Photometric Science Alert system \citep{Hodgkin2021}. There are many plausible mechanisms that could be responsible for the light variations observed in almost every single PMS star (see e.g. \citealt{Siwak2019, Nagy2021, Fischer2022}), but the most exciting are those resulting from substantially enhanced accretion. 

Up until the time of writing, two outbursts observed by Gaia have been spectroscopically confirmed as FUors: Gaia17bpi \citep{Hillenbrand2018_ApJ869146H} and Gaia18dvy \citep{SzegediElek2020_ApJ899130S}, and four as EXors: Gaia18dvz (ESO-H${\alpha}$99 \citealt{Hodapp2019_AJ158241H}); Gaia20eae (originally found in ground-based data as PGIR20dwf by \citealt{hankins2020}) and studied in detail by \cite{Ghosh2021} and \cite{Fernando2022}; Gaia19fct \citep{Park2022,Miller2015}; and Gaia22dtk \citep{Kuhn2022}. Two other eruptive young stars identified from Gaia variability do not clearly fall either into the FUor or the EXor category but share properties of both; these are Gaia19ajj \citep{Hillenbrand2019_AJ158240H} and Gaia19bey \citep{Hodapp2020_AJ160164H}. In addition, the Gaia Alert system pointed our attention to the temporary fading in ESO-H$\alpha$~148 (since then also known as Gaia21elv), first analysed by \citet{Contreras-Pena2019}, but only now confirmed as a bonafide FUor \citep{Nagy2023}.

The object of interest here, Gaia21bty, was identified by the Gaia Science Alert system as a considerable 2.9~mag brightening of a very reddened star, 2MASS 17251419-3708140 or Gaia DR3 5974356442830973056. The source is located about a degree to the west of $\upsilon$ Scorpii, in the Galactic mid-plane. 
Based on a machine-learning classification using WISE and Gaia DR2 data, \citet{Marton2016} gave a high probability (89.22\%) that this is a young stellar object (YSO).
The star was also listed among YSO candidates by \citet{Kuhn2021}. 
An initial analysis of the Gaia21bty spectrum, taken near its maximum brightness, indicated that it may be an FU~Ori-type star \citep{Hillenbrand2021}. 

Here we present and analyze new and archival data for Gaia21bty and draw conclusions about its nature. In Section~\ref{sec:observations}, we describe the reduction of the archival photometry, and our new spectroscopic and photometric data obtained during the outburst until early-2023. Then, in Section~\ref{sec:properties}, we estimate spectral class, distance, and briefly characterise the environment of Gaia21bty. Results obtained from the analysis are presented and discussed in Section~\ref{sec:analysis}, and then summarised in Section~\ref{sec:summary}.

\begin{figure*}
    \centering
    \includegraphics[width=2.\columnwidth]{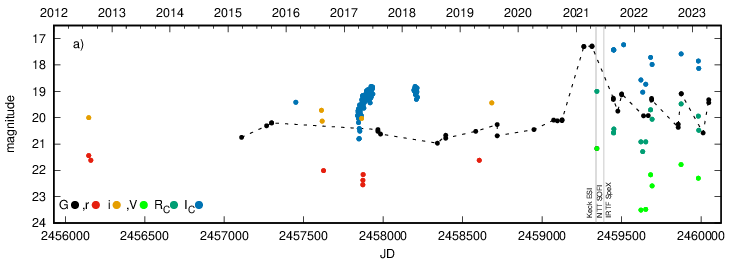}\\
    \includegraphics[width=2.\columnwidth]{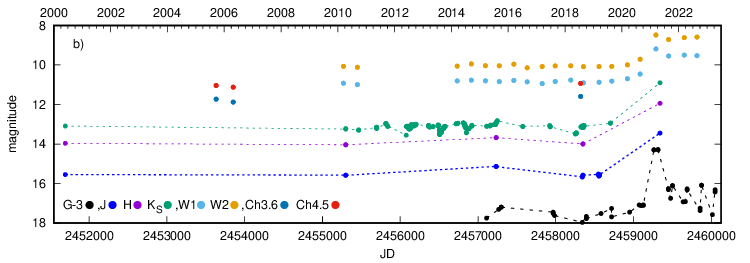}
    \caption{Optical (top) and infrared (bottom) light curves of Gaia21bty. Moments of spectroscopic observations are indicated in the first plot.
    $JHK_S$ pre-outburst observations were obtained by 2MASS and VVVX surveys, while NTT equipped with SOFI covered the maximum stage. Due to the nebulosity, the brightness in the WISE bands $W1$ (3.4 $\mu$m) and $W2$ (4.6 $\mu$m) is higher by about 1~mag than in the respective SPITZER 3.6~$\mu$m and 4.5~$\mu$m channels from 2018. Gaia light curve (shifted by 3 mag) is plotted again to facilitate the navigation. The error bars are comparable in size or smaller than the points, and are omitted for clarity. The comas in the legends are to separate photometric systems.}
    \label{fig:lc}
\end{figure*}

\begin{figure}
    \centering
    \includegraphics[width=1.\columnwidth]{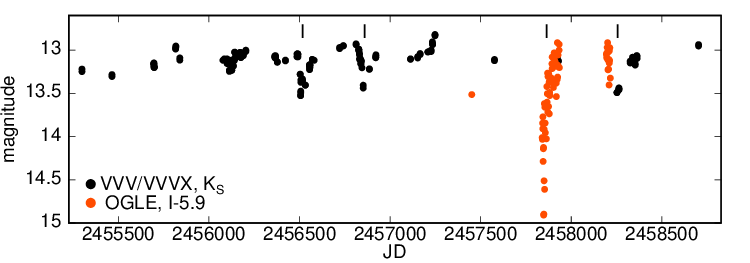}
    \caption{Enhanced view of the information in Figure~\ref{fig:lc} showing the dips in $I$ and $K_S$ bands during the pre-outburst stage.}
    \label{fig:dips}
\end{figure}

\begin{table}
\centering
\caption{Pre-outburst optical and infrared photometry of Gaia21bty extracted from public-domain, calibrated images, gathered by different surveys.}
\label{tab:photometry_ours_pre}
\begin{tabular}{ccccc}
\hline \hline
JD             & Filter& Mag    & Unc  & Survey \\ \hline
2456147.63402  & $r$   & 21.44  & 0.08 & VPHAS+ \\
2456147.64006  & $i$   & 20.00  & 0.05 & VPHAS+ \\
2456160.53970  & $r$   & 21.62  & 0.10 & VPHAS+ \\
2457872.72827  & $g$   & 24.87  & 0.92 & DECaPS \\
2457624.67796  & $r$   & 22.01  & 0.17 & DECaPS \\
2457871.83275  & $r$   & 22.55  & 0.20 & DECaPS \\
2457872.72721  & $r$   & 22.38  & 0.18 & DECaPS \\
2457873.82431  & $r$   & 22.16  & 0.18 & DECaPS \\
2458604.83248  & $r$   & 21.62  & 0.13 & DECaPS \\
2457611.66478  & $i$   & 19.72  & 0.15 & DECaPS \\
2457616.69413  & $i$   & 20.13  & 0.16 & DECaPS \\
2457864.82449  & $i$   & 20.03  & 0.16 & DECaPS \\
2458683.60124  & $i$   & 19.44  & 0.15 & DECaPS \\
2457611.66545  & $z$   & 17.67  & 0.13 & DECaPS \\
2457616.69345  & $z$   & 17.98  & 0.13 & DECaPS \\
2457864.82518  & $z$   & 18.00  & 0.12 & DECaPS \\
2458683.60192  & $z$   & 17.54  & 0.11 & DECaPS \\
2457611.66612  & $Y$   & 17.14  & 0.11 & DECaPS \\ 
2457616.69277  & $Y$   & 17.39  & 0.12 & DECaPS \\
2457864.82587  & $Y$   & 17.36  & 0.12 & DECaPS \\ 
2458683.60262  & $Y$   & 16.97  & 0.11 & DECaPS \\ \hline
2453632.54274  & [8.0] & 10.461 & 0.172 & GLIMPSE \\
2453632.54282  & [4.5] & 11.038 & 0.058 & GLIMPSE \\  
2453632.57065  & [3.6] & 11.731 & 0.053 & GLIMPSE \\
2453632.57065  & [5.8] & 10.823 & 0.060 & GLIMPSE \\
2453852.58020  & [3.6] & 11.879 & 0.085 & GLIMPSE \\
2453852.58020  & [5.8] & 10.669 & 0.348 & GLIMPSE \\
2453852.59760  & [4.5] & 11.121 & 0.085 & GLIMPSE \\
2453852.59760  & [8.0] & 10.395 & 0.278 & GLIMPSE \\ 
2458319.78485  & [3.6] & 11.591 & 0.026 & GLIMPSE \\
2458319.79077  & [4.5] & 10.933 & 0.031 & GLIMPSE \\ \hline
\end{tabular}
\end{table}

\begin{table}
\centering
\caption{Outburst (top part) and post-outburst photometry of Gaia21bty 
extracted from images gathered during this work.}
\label{tab:photometry_ours_post}
\begin{tabular}{ccccc}
\hline \hline
JD             & Filter& Mag    & Unc  & Telescope/Instrument \\ \hline
2459340.68850  & $J$   & 13.449  & 0.044 & NTT/SOFI   \\ 
2459340.69101  & $H$   & 11.937  & 0.029 & NTT/SOFI   \\
2459340.69352  & $K_S$ & 10.902  & 0.028 & NTT/SOFI   \\
2459344.67628  & $V$   & 21.17  & 0.12 & NTT/EFOSC2 \\
2459344.66854  & $R_C$ & 19.00  & 0.05 & NTT/EFOSC2 \\ \hline
2459449.23811  & $R_C$ & 20.57  & 0.32 & Elisabeth/SHOC  \\
2459449.24054  & $I_C$ & 17.43  & 0.13 & Elisabeth/SHOC  \\
2459450.22208  & $R_C$ & 20.58  & 0.15 & Elisabeth/SHOC  \\
2459450.21666  & $I_C$ & 17.42  & 0.13 & Elisabeth/SHOC  \\
2459451.23182  & $R_C$ & 20.43  & 0.11 & Elisabeth/SHOC  \\
2459451.22910  & $I_C$ & 17.43  & 0.13 & Elisabeth/SHOC  \\
2459513.24948  & $I_C$ & 17.23  & 0.16 & Elisabeth/SHOC  \\
2459621.86600  & $V$   & 23.51  & 0.23 & Danish/DFOSC \\
2459621.87700  & $R_C$ & 20.92  & 0.08 & Danish/DFOSC \\
2459621.88200  & $I_C$ & 18.57  & 0.08 & Danish/DFOSC \\
2459632.76602  & $R_C$ & 21.29  & 0.15 & Danish/DFOSC \\
2459632.77005  & $I_C$ & 19.03  & 0.11 & Danish/DFOSC \\
2459651.85932  & $V$   & 23.48  & 0.27 & Danish/DFOSC \\
2459651.82779  & $R_C$ & 20.92  & 0.11 & Danish/DFOSC \\
2459651.82959  & $I_C$ & 18.73  & 0.09 & Danish/DFOSC \\
2459682.86387  & $V$   & 22.17  & 0.14 & Danish/DFOSC \\
2459682.87036  & $R_C$ & 19.70  & 0.07 & Danish/DFOSC \\
2459682.90521  & $I_C$ & 17.71  & 0.08 & Danish/DFOSC \\
2459691.89099  & $V$   & 22.59  & 0.25 & Danish/DFOSC \\
2459691.88942  & $R_C$ & 20.24  & 0.08 & Danish/DFOSC \\
2459691.90696  & $I_C$ & 17.98  & 0.08 & Danish/DFOSC \\
2459874.49908  & $V$   & 21.78  & 0.14 & Danish/DFOSC \\
2459874.51736  & $R_C$ & 19.48  & 0.06 & Danish/DFOSC \\
2459874.53502  & $I_C$ & 17.58  & 0.08 & Danish/DFOSC \\
2459982.86807  & $V$   & 22.30  & 0.15 & Danish/DFOSC \\
2459982.83635  & $R_C$ & 19.94  & 0.12 & Danish/DFOSC \\
2459982.85189  & $I_C$ & 17.85  & 0.09 & Danish/DFOSC \\
2459985.87106  & $R_C$ & 20.48  & 0.15 & Danish/DFOSC \\
2459985.85622  & $I_C$ & 18.13  & 0.07 & Danish/DFOSC \\
\hline
\end{tabular}
\end{table}

\section{Observations and the data reduction}
\label{sec:observations}

\subsection{Optical and infrared photometry}
\label{sec:phot}
In order to characterise Gaia21bty in the pre-outburst stage, we collected archival optical and infrared photometry from several sources (Sec.~\ref{sec:phot_arch}). Then, we obtained new photometry to characterise the outburst and post-outburst stages (Sec.~\ref{sec:phot_new}).

\subsubsection{Archival photometry}
\label{sec:phot_arch}

The earliest optical data were obtained on two close epochs, 2012 August 8 and 21, by the VST Photometric H${\alpha}$ Survey of the Southern Galactic Plane and Bulge (VPHAS+; \citealt{Drew2014}). The survey was conducted by means of the 2.6-m ESO VLT Survey Telescope (VST) at Cerro Paranal, equipped with the OMEGACAM and $ugriH{\alpha}$ filters. We collected calibrated fits images from the public ESO archive and unambiguously detected our source only in the $ri$-filters. The star was only marginally detectable in the $H{\alpha}$ filter. We extracted the photometry using {\sc daophot} procedures \citep{Stetson1987} distributed within the \texttt{astro-idl} library. To ensure optimal extraction, an aperture of 9 pixel (1.92~arcsec) was used for the target and four comparison stars. The inner and outer sky annuli were specifically adjusted for each star to enable accurate sky level calculation. 
We took the $ri$ magnitudes of the comparison stars from the APASS9 catalogue \citep{henden2015,henden2016}. The four comparison stars available on the same CCD chip as our target have bluer colour indices, but careful extrapolation enabled us to obtain approximate transformation to the Sloan system. We plot these results in Figure~\ref{fig:lc}a and show in Table~\ref{tab:photometry_ours_pre}. In the same figure (and also in Fig.~\ref{fig:lc}b with infrared data) we present Gaia $G$-band data, connected with lines to facilitate navigation.

Optical data were also gathered during a few epochs between 2016 August 11 -- 2019 July 19 by the DECam Plane Survey (DECaPS; \citealt{Schlafly2018}). The survey was conducted by means of the Dark Energy Camera (DECam) equipped with the $grizY$ filters, installed on the 4-m Blanco telescope at the Cerro Tololo Inter-American Observatory (CTIO) in Chile. We downloaded calibrated fits images from the survey's public archive and (for consistency with VPHAS+ and our future observations) extracted the photometry using the same apertures (in arcseconds) as for the VPHAS+ data.
The $gri$ magnitudes of the three-to-five (depending on saturation) comparison stars were taken from the APASS9 catalogue. To obtain the $z$ and $Y$ magnitudes, we plotted their broad-band SED using their APASS9 and 2MASS magnitudes \citep{cutri2003}, and then used spline interpolation for the filters' effective wavelengths (Fig.~\ref{fig:lc}a, Tab.~\ref{tab:photometry_ours_pre}). Note that Gaia21bty is marginally detectable in the $g$-band (with 0.9~mag uncertainty), but only during one night.

Between 2016 March 2 -- 2018 April 7, the star was observed by the Optical Gravitational Lensing Experiment (OGLE; \citealt{Udalski2015}), conducted with the 1.3-m Warsaw University telescope at the Las Campanas Observatory in Chile. The data were obtained using the third generation OGLE-IV mosaic camera equipped with the interferometric $VI$ filters manufactured by Asahi Spectra, but significant detection of Gaia21bty was possible only for the $I$-band ($\lambda_{\rm eff}=793.2$~nm). The photometry used in this paper (Fig.~\ref{fig:lc}a, Fig.~\ref{fig:dips}) was extracted by means of the dedicated OGLE pipeline, as described in \citet{Udalski2015}.

The first pre-outburst infrared measurements in $JHK_S$ filters used in this paper were obtained by the 2MASS survey on 2000 May 29 \citep{Skrutskie2006}. Then, between 2010 April 11 -- 2019 August 6, the star was observed from the subsidiary summit of Cerro Paranal by the 4-m Visible and Infrared Survey Telescope for Astronomy (VISTA) equipped with the near-infrared imaging camera (VISTA IR Camera, VIRCAM) and $ZYJHK_S$ filters \citep{Sutherland2015}. The data were obtained within the public ESO near-infrared VISTA Variables in the Via Lactea (VVV) survey and its extension (VVVX), which scanned the Milky Way bulge and the section of the mid-plane where star formation rate is supposed to be high \citep{Minniti2010}. 
The photometry (Fig.~\ref{fig:lc}b, Fig.~\ref{fig:dips}) was extracted from a pre-release version of version 2 of the VVV Infrared Astrometric Catalogue (VIRAC2, see \citet{Smith2018}, and Smith et al., in prep). This provides time series photometry and five parameter absolute astrometric fits for sources using VVV/VVVX observations from 2010 to 2019. Individual sources were extracted by profile fitting photometry using DoPHOT \citep{Schechter1993, Alonso-Garcia2018}, with improved photometric calibration. The astrometry is calibrated on to the Gaia DR3 reference frame. 

To characterise the progenitor at longer wavelengths, we analysed data from the {\sl Spitzer} \citep{Werner2004} and the Wide-field Infrared Survey Explorer \citep[WISE;][]{Wright2010} space telescopes. In the Spitzer Heritage Archive we found five observations that covered Gaia21bty obtained with the Infrared Array Camera \citep[IRAC;][]{Fazio2004}, all as part of larger mapping projects. The first four measurements were performed during the cryogenic phase in 2005 and 2006. In two cases, Gaia21bty was observed at 3.6 and 5.8\,{\micron}, while in two other cases at 4.5 and 8\,{\micron}. The data corresponding to the latest epoch were obtained only in the 3.6 and 4.5\,{\micron} bands about 12\,years later, during the post-cryogenic (warm) phase. For photometry, we used the corrected basic calibrated data (CBCD) frames generated by the IRAC pipeline, all downloaded from the Spitzer Science Center. Aperture photometry was used to extract flux densities. The aperture radius was set to 2 pixels ($\sim$2\farcs4), while the background was estimated in an annulus between radii of 2 and 6 pixels. The centroid of the source was determined by employing a first moment box centroider algorithm\footnote{\url{https://irsa.ipac.caltech.edu/data/SPITZER/docs/dataanalysistools/tools/contributed/irac/box\_centroider/}}. In addition to aperture correction, array-location and pixel phase correction were also performed. For the latter two corrections, the \texttt{irac aphot corr} routine\footnote{\url{https://irsa.ipac.caltech.edu/data/SPITZER/docs/dataanalysistools/tools/contributed/irac/iracaphotcorr/}} was used, while the aperture correction factors were taken from the IRAC Instrument Handbook~v3.01\footnote{\url{https://irsa.ipac.caltech.edu/data/SPITZER/docs/irac/iracinstrumenthandbook/}}. Except for one measurement where the target was observed only on one frame, we have multiple data points corresponding to the 2--6 frames available for the source. In the latter cases the final flux densities and their uncertainties were obtained as the mean and the error 
of the mean of the individual values, respectively. The photometric data are shown in Table~\ref{tab:photometry_ours_pre} and plotted in Figure~\ref{fig:lc}b. The quoted uncertainties are the quadratic sums of measurement and calibration errors, where the latter is adopted to be 2\%. 

Gaia21bty was also observed with the Multiband Imaging Photometer for {\sl Spitzer} \citep[MIPS;][]{Rieke2004ApJS..154...25R}. However, strong contamination from the nearby IR-nebula prevented us from obtaining accurate measurements in all MIPS bands, only a 3\,$\sigma$ upper limit of 0.15\,Jy could be estimated at 24\,{\micron}. This will have some repercussions when trying to classify our target (Sec.~\ref{sec:sed}).

Two epochs, observed six months apart, were also collected during the fully cryogenic phase of WISE in $W1$, $W2$, $W3$ and $W4$ bands, centered at 3.4, 4.6, 12, and 22~$\mu$m respectively. However, we noted that as in the case of Spitzer, no distinct point source is visible on top of the surrounding nebula in the $W3$ and $W4$ images, therefore, we rejected these catalogue measurements from further analyses. The NEOWISE mission continued to operate in $W1$ and $W2$-bands \citep{Mainzer2011} providing long-term light curves covering the outburst. 
Starting from 2014, usually over a dozen individual measurements per filter were obtained during each of two (typically 2 days long) pointings towards our target. We downloaded individual catalogue measurements, removed obvious outliers, and computed averages and their standard deviations per each epoch. In spite of the nebulosity having impact on the real brightness level also in $W1$ and $W2$ bands, this is mostly a systematic effect, therefore the light curves still carry valuable information about the mid-infrared brightness evolution.

\subsubsection{New photometry}
\label{sec:phot_new}

To characterise the outburst itself and the post-outburst behavior, on 2021 May 10, we used the 3.58-m New Technology Telescope (NTT) of the European Southern Observatory (ESO) in La Silla, Chile. The ESO Faint Object Spectrograph and Camera version 2 (EFOSC2) camera equipped with $BVR_C$ filters was used, but we detected our star in the $V$- and $R_C$-filter images only; The target remains undetected both on the three individual 40~sec long, and on the combined (120~sec in total) $B$-filter exposures. 
Then, in August and October, 2021, we used the 1-m {\it Elizabeth} telescope located at the South African Astronomical Observatory (SAAO). The telescope was equipped with the Sutherland High Speed Optical Camera (SHOC).
We observed the star in the $VR_CI_C$ filters, but we got no detection in the $V$ band. This is due to the fact that contrary to our expectations, the source faded considerably by this time.

In February 2022 -- February 2023 we observed the star on eight different nights using the Danish 1.54-m telescope at La Silla, Chile. 
We used the Danish Faint Object Spectrograph and Camera (DFOSC) equipped with Johnson-Cousins filters. 
The star remained undetected in the $B$ filter either in any single 300~sec individual exposures, or in the stacked (in total 30-50~min, depending on night) images. 
Positive detection was almost always achieved for the $V$-band: the star remained undetected only during two nights with poor seeing conditions. The target was always well visible in the $R_C$- and $I_C$-band images.

We corrected our new raw images for bias, dark (when necessary), and flatfield using {\sc iraf}, and then applied {\sc daophot} procedures.
The magnitudes of the suitable comparison stars, four in the $4.1\times4.1$~arcmin field of view (FoV) on NTT, three in the $2.85\times2.85$~arcmin FoV on {\it Elizabeth}, and 34 in the $13.7\times13.7$~arcmin FoV of the Danish telescope, were obtained from the APASS9 and 2MASS catalogues \citep{henden2015,henden2016, cutri2003}. The missing $R_C$ and $I_C$ magnitudes were obtained by SED interpolation, like before. 
We used the same aperture size (in arcseconds) for the target and the (usually common) comparisons stars as for pre-outburst VPHAS+ and DECaPS data. We corrected our results for the colour extinction assuming the second-order colour extinction coefficients $\beta=-0.008$ both for $R_C$ and $I_C$ bands and $-0.01$ for the $V$-band \citep{Siwak2018a,Siwak2018b}. The small FoV of NTT/EFOSC2 and {\it Elisabeth}/SHOC cameras prevented us from observing comparison stars having colour indices similar to Gaia21bty to enable proper colour transformation, but we estimate that the lack of the full transformation to the standard system should not lead to systematic error higher than 0.15~mag. We plot the results in Figure~\ref{fig:lc}a and list the obtained magnitudes in Table~\ref{tab:photometry_ours_post}. 

On 2021 May 6 (at the maximum brightness) we obtained images in $JHK_S$ filters with the infrared spectrograph and imaging camera Son of ISAAC (SOFI, \citealt{Sofi1998Msngr..91....9M}), operating at the Nasmyth focus of NTT. The photometry was obtained using 20 comparison stars. They were carefully selected from the 2MASS catalogue \citep{cutri2003} bearing in mind both brightness and colour index similar to those of Gaia21bty. As the images were obtained in single position only and no other calibration images were available, we found and removed the {\it magnitude-coordinate} dependency by means of a two-dimensional quadratic function fit. No further dependency on colour index was found. We list the final results in Table~\ref{tab:photometry_ours_post} and plot in Figure~\ref{fig:lc}b. As expected, the NEOWISE mission continued as the outburst proceeded and contributed to post-outburst mid-infrared light curve as well.

\begin{figure*}
    \centering
    \includegraphics[width=2.\columnwidth]{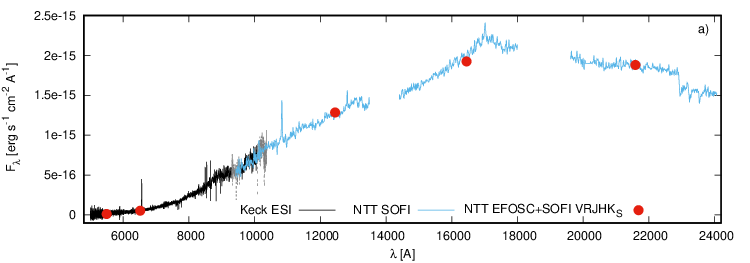}
    \includegraphics[width=2.\columnwidth]{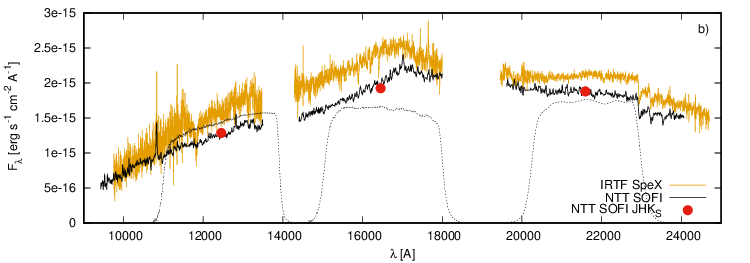}
    \includegraphics[width=2.\columnwidth]{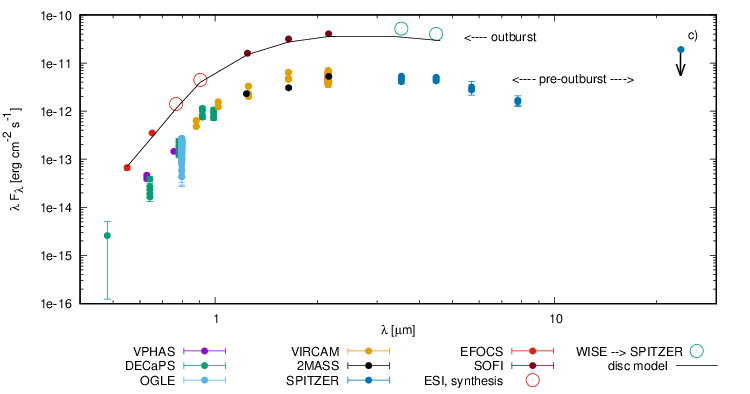}
    \caption{Optical (a) and infrared (a,b) telluric-corrected spectra, scaled to fluxes.
    The noisy part below 5000~\AA, and regions heavily contaminated by telluric lines, were marked by grey or removed. The profiles of SOFI's $JHK_S$ filters are plotted in panel b with dots for auxiliary purposes. Panel c shows the pre-outburst and the ouburst SED of Gaia21bty. The Sloan $iz$ magnitudes are synthetic, obtained from the optical spectrum integration. The NEOWISE $W1$ and $W2$ data from 2021 March 13 are transformed to Spitzer 3.6 and 4.5~$\mu m$ bands, and indicated by large open circles to emphasize their larger beam. Result from our disc modelling is plotted as the dark line.}
    \label{fig_sed}
\end{figure*}

\subsection{Optical and infrared spectroscopy}
\label{sec:spectra_new}

Keck~II equipped with the high throughput Echellette Spectrograph and Imager (ESI, \citealt{ESI2002PASP..114..851S}) was used to obtain the optical spectrum of Gaia21bty on 2021 May 1. ESI provided complete coverage from 390 to 1030~nm in a single 1800~sec exposure. We used the cross-dispersed spectroscopy mode along the $0.5\times20$~arcsec slit, providing a resolving power up to $R=8000$ (37~km~s$^{-1}$). The spectra were processed in a standard way for bias and flatfield, and then wavelength-calibrated and heliocentric velocity-corrected by means of the dedicated {\sc Makee} reduction package, written by Tom Barlow. As no telluric standard was observed, we removed the telluric lines using the {\sc molecfit} software \citep{Smette2015, Kausch2015}. We then scaled the observed spectrum to the absolute fluxes using nearly-simultaneous visual and infrared photometry obtained at La Silla (Tab.~\ref{tab:photometry_ours_post}). We show the final result in Figure~\ref{fig_sed}a.  

Six days later, on 2021 May 6, we obtained the first near-infrared spectrum by means of SOFI on NTT. We used the low resolution blue and red grisms with GBF and GRF order sorting filters through the 0.6~arcsec wide slit, covering $0.95-1.64$ and $1.53-2.52$~$\mu$m with $R=930$ and 980, respectively. We reduced the spectra in {\sc iraf} according to the recipe given in the SOFI manual: first the sky level was approximately removed by subtraction of images obtained in A and B nod positions, and then the sky residuals were carefully removed during the spectrum extraction with the \texttt{apall} task. The Xenon lamp spectra and respective line lists provided for SOFI users were used for the wavelength calibration performed with the \texttt{(re)identify} and \texttt{dispcor} tasks. Then we removed atmospheric lines by means of the \texttt{telluric} task, in which we interactively scaled the normalised spectrum of the telluric standard star HD74194, obtained on similar airmass but 3 hours earlier than the target. Finally we combined the cleaned images to form average 'blue' ($0.95-1.64$~$\mu$m) and 'red' ($1.53-2.52$~$\mu$m) spectra (Fig~\ref{fig_sed}a, b).

The second near-infrared spectrum was obtained on 2021 June 23 (i.e., 48 days after SOFI observations) by means of the NASA Infrared Telescope Facility (IRTF) equipped with SpeX \citep{SpeX2003PASP..115..362R}. SpeX is a medium-resolution infrared spectrograph, which we used in the Short XD (SXD) mode, providing $0.7-2.55$~$\mu$m spectra with $R=2000$ through the $0.3\times15$~arcsec slit. The total integration time of six dithered spectra was 717~sec. After the spectrum extraction by the dedicated pipeline, telluric correction was performed by division of the Gaia21bty spectrum by the spectrum of A0 star HIP~86098, located 2~deg from and observed immediately after the target at exactly the same 1.84 airmass. Then, the heliocentric velocity correction was applied. Both the SOFI and the SpeX spectra are plotted in Fig.~\ref{fig_sed}b. One can note slightly higher flux obtained by SpeX than by SOFI.
Due to the lack of simultaneous $JHK_S$ photometry during the IRTF/SpeX observations and strong variability of the source, we can not exclude that the enhanced flux seen by IRTF/SpeX relative to NTT/SOFI is real.

\begin{figure}
    \centering
    \includegraphics[width=1\columnwidth]{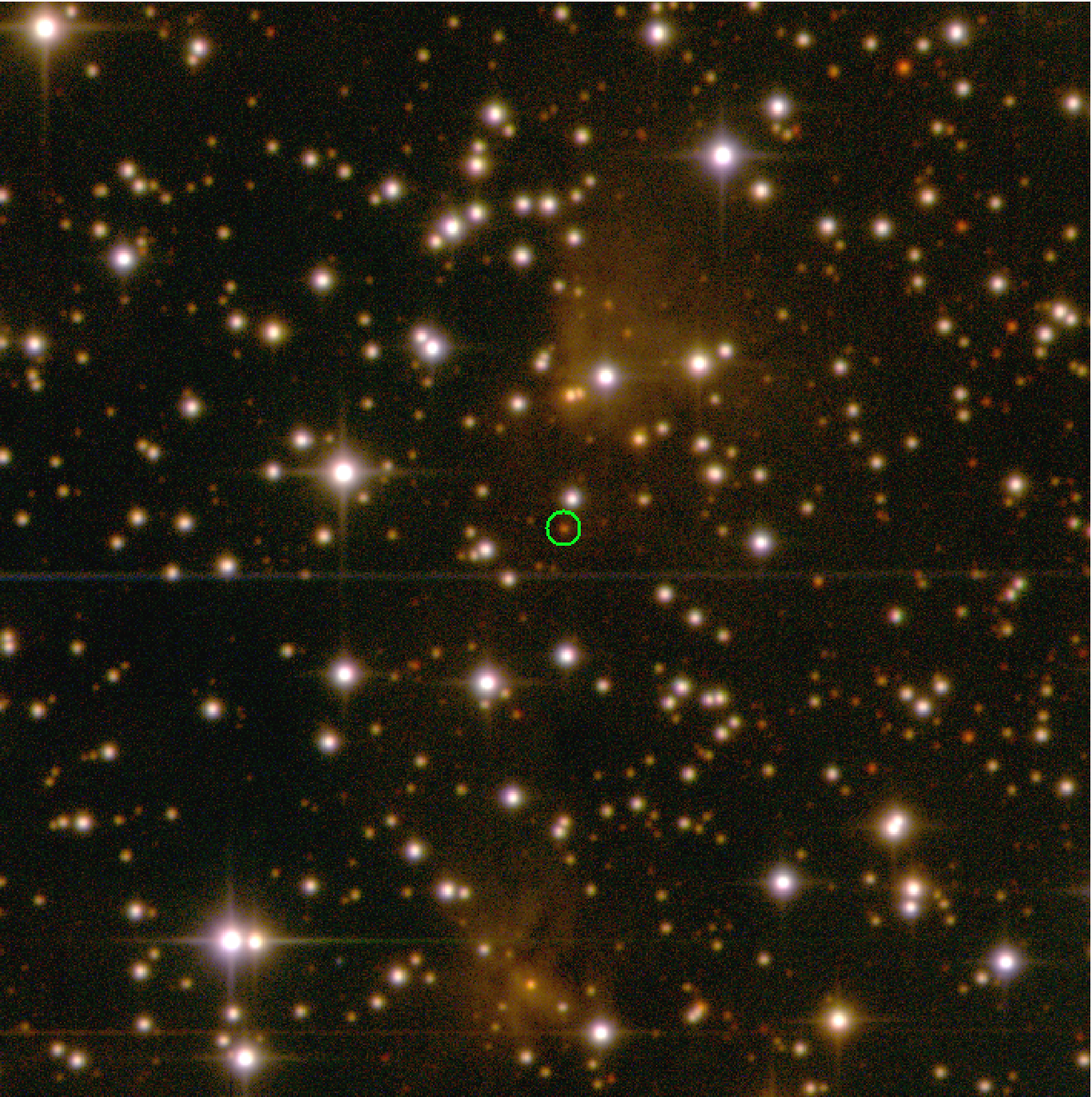}
    \caption{Composite $BVR$ $4.5\times4.5$~arcmin image, obtained on 2022 April 12 by 1.54-m Danish telescope at La Silla. Gaia21bty, having then a brightness of $V=22.17$~mag and $R_C=19.70$~mag, is encircled with a 4~arcsec radius aperture. North is up, east is to the left.}
    \label{fig_picture}
\end{figure}

\begin{figure}
    \centering
    \includegraphics[width=1\columnwidth]{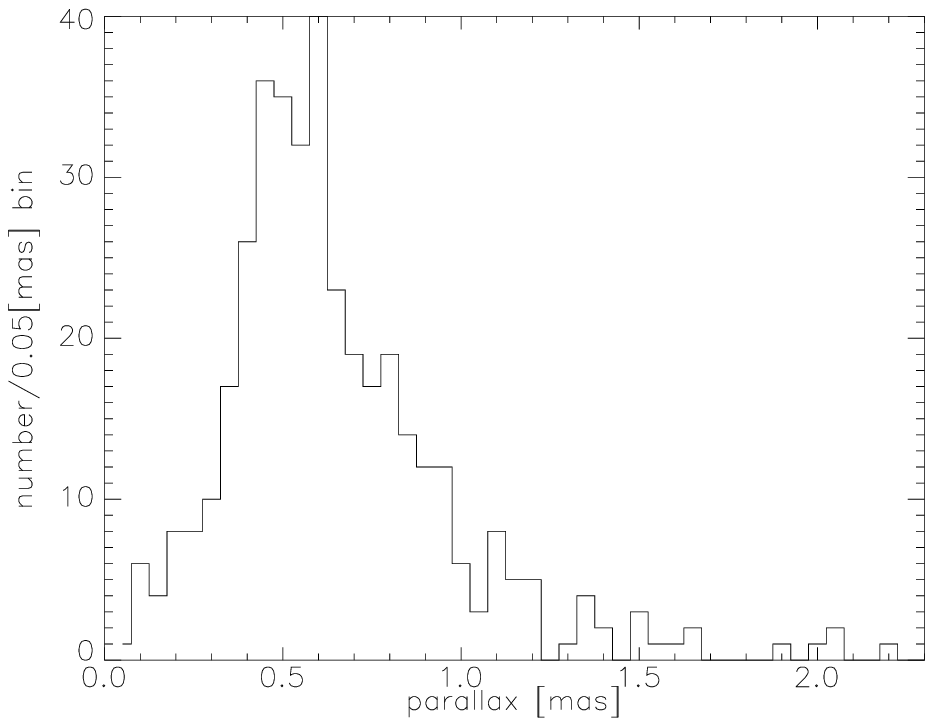}
    \caption{Distribution of parallaxes of red objects in the 4$'$ vicinity of Gaia21bty.}
    \label{fig_distances}
\end{figure}

\begin{figure}
    \centering
    \includegraphics[width=1\columnwidth]{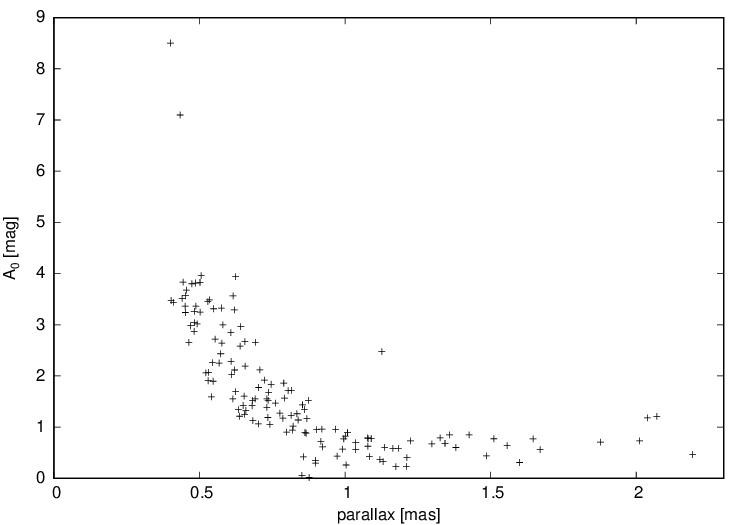}
    \caption{Interstellar extinction $A_0$ vs. parallax, as obtained from Gaia DR3 for stars with well-determined parallaxes.}
    \label{fig_ext_parallax}
\end{figure}

\section{Properties of Gaia21bty}
\label{sec:properties}

\subsection{Spectral energy distribution and classification of progenitor}
\label{sec:sed}

We constructed the spectral energy distribution (SED) of Gaia21bty from the archival and new data (Sec.~\ref{sec:phot}) obtained prior and during the outburst, as shown in Figure~\ref{fig_sed}c. Appropriate zero points, necessary to transform observed magnitudes to flux units, were taken from the Spanish Virtual Observatory (SVO) filter profile database\footnote{\url{http://svo2.cab.inta-csic.es/theory/fps/}}. 

The pre-outburst SED is constructed from all data but Gaia, taken before 2020. Thus, some of them, like the OGLE $I$- and VVV/VVVX $K_S$-filter data, allow one to visualize the spread of energy received in the optical bands during the phenomena unrelated to the outburst, i.e., the obscuration events (Sec.~\ref{sec:var_pre}). As mentioned in Section~\ref{sec:phot}, due to the bright nebula and 5.4$\times$ larger pixel scale, the NEOWISE data (6.5~arcsec~pix$^{-1}$) indicate a much higher brightness of the target than those obtained by Spitzer (1.2~arcsec~pix$^{-1}$) at similar wavelengths. For this reason the NEOWISE measurements are omitted from the pre-outburst SED.

The spectral index $\alpha$ \citep{Lada1987} estimated between the Spitzer 4.5 and (the upper limit of) 24 $\mu$m fluxes (equation 7 in \citealt{Kuhn2021}) is $\leq$0.876, which suggests Class~I, flat or Class~II evolutionary stage. 
However, Figure~\ref{fig_sed}c shows steady negative spectral index between 4.5, 5.8, and 8~$\mu$m, and not a deep silicate absorption at 8~$\mu$m. In these circumstances, the spectral index calculated from equation~9 in their paper, utilizing 8.0~$\mu$m magnitude, results in $\alpha=-1.7$.
In addition, Figure~\ref{fig_ccd} suggests quite low near-infrared excess fluxes above the main sequence.
All the above supports Class~II classificiation for our target.

The outburst fluxes were measured by NTT within 4 days, on 2021 May 6 in the $JHK_S$, and on May 10 in the $VR_C$ filters. At that time the star was still in the maximum brightness, or just started the fade. Despite this uncertainty, throughout this paper we consequently assume that these data represent the maximum stage.

In addition to our data, thanks to the similarity between Spitzer 3.6 \& 4.5~$\mu m$ and NEOWISE $W1$ \& $W2$ bands, we estimated the influence of the nebulosity on the latter photometry. For this purpose we used the nearly-simultaneous measurements obtained in August 2018 (Fig.~\ref{fig:lc}b). First, we converted the $W1$ and $W2$ magnitudes to fluxes in Jansky unit, then we aligned these light curves to match the respective Spitzer's data, and converted back to magnitudes. The nebulosity-corrected NEOWISE measurements obtained on 2021 March 13 (certainly the outburst maximum) are $W1=9.318\pm0.031$ and $W2=8.619\pm0.033$~mag. We mark these points as open circles in Figure~\ref{fig_sed}c to indicate the indirect nature of these measurements. The same applies to the $i$- and $z$-band Sloan points, obtained from spectral synthesis of the Keck/ESI spectrum.

We will return to the pre-outburst and outburst SEDs analysis in Sections~\ref{sec:progenitor} and ~\ref{sec:sed_mod}, after the interstellar extinction estimation.

\subsection{Location and distance estimate}
\label{sec:location}

Gaia21bty  ($\alpha_{\rm J2000}$ = 17$^{\rm h}$ 25$^{\rm m}$ 14$\fs$19, $\delta_{\rm J2000}$ = $-$37$^{\circ}$ 08$'$ 14$\farcs$17) 
is located in an optically-dark region very close to the Milky Way mid-plane ($l=350.79501^{\circ}$, $b=-0.84115^{\circ}$), in the southern part of Scorpius. This region was targeted by the Spitzer GLIMPSE~II project, one of the goals of which was identification of mid-infrared dust bubbles formed by hot O and B type young stars in massive star forming regions, to trace recent and current star formation in the Milky Way \citep{Churchwell2007}. Yet, the MWP1G350770-008600S bubble adjacent to the dark region from which Gaia21bty emerged, was identified only later via the online citizen science website ‘The Milky Way Project’ \citep{Simpson2012}. 

In the pre outburst stage, Gaia21bty was a relatively faint ($G=20.364\pm0.027$~mag) and strongly reddened object (Sec.~\ref{sec:phot}, Tab.~\ref{tab:photometry_ours_pre}), therefore there is no parallax in the third Gaia Data Release (DR3, \citealt{GaiaDR3}), nor in the earlier data releases. Gaia21bty has no proper motion in the Gaia~DR3 catalogue. Therefore we examined the comoving stars in the VIRAC2 catalogue. The proper motion of Gaia21bty in right ascension and declination was measured there as $pmra=0.34\pm0.24$~mas~yr$^{-1}$ and $pmdec=-0.54\pm0.28$~mas~yr$^{-1}$, respectively. Converting to Galactic coordinates, the proper motion in the Galactic longitude is $pml = 0.03\pm0.25$~mas~yr$^{-1}$ and in the Galactic latitude is $pmb = -0.64\pm0.27$~mas~yr$^{-1}$. The essentially zero motion in the longitude direction indicates that it is probably within 5~kpc. This constrain is insufficient for our purposes.  We thus searched for known YSOs in the vicinity of Gaia21bty with determined parallaxes. 

Using the list of \citet{Kuhn2021}, we found a dozen Class II YSOs in a 10$'$ radius of Gaia21bty, out of which only one, SPICY 53729, has well-determined parallax $0.4847\pm0.0636$~mas, resulting in the distance $d\approx 2.06$~kpc.
We also examined properties of YSO candidates from \cite{Marton2016, Marton2019} within the same 10$'$ radius around Gaia21bty. Only three Class~II objects with well determined parallaxes were found in their first study, and the mean value is $0.49\pm0.03$~mas (2.04~kpc). In the 2019 catalogue, we found 74 stars for which the probability of being YSO is higher than 70\%. We removed the stars with negative parallaxes and considered only those having renormalised unit weight error $RUWE \leq 1.4$ and parallaxes at least five times their errors.
Based on 13 stars meeting these criteria, we obtained the median parallax $0.59\pm0.20$~mas, which corresponds to $d=1.70^{+0.87}_{-0.43}$~kpc. This is in line with the VLBA results of \citet{Wu2014}, who found that the typical distances to all nearby SFRs from the Saggitarius arm are $\leq 2$~kpc (see the first few lines in Tab.~3 of their paper).
 
Furthermore, we checked whether red stars seen towards Gaia21bty could be members of Galactic spiral arms along the line of sight, 
specifically looking for those associated with the star forming region related to the aforementioned bubble.
Based again on the Gaia DR3 catalogue and considering only stars with $RUWE \leq 1.4$, but now adding a selection for red colour index ($1.0 \leq B_P-R_P \leq 4.5$~mag), we analysed the parallaxes of objects localised within 4$'$ radius around Gaia21bty. The distribution of parallaxes in  0.05~mas bins (Figure~\ref{fig_distances}) appears to be double-peaked with peaks at 0.6~mas (the formal median value of unbinned parallaxes equals to 0.588~mas, similarly as obtained above based on \citealt{Marton2019}), and at 0.45~mas. 

Compiling our results obtained for YSOs in the same direction as Gaia21bty \citep{Wu2014, Kuhn2021, Marton2016, Marton2019}, we conclude that the distance to Gaia21bty is probably close to 1.7~kpc. It is certainly not shorter than 1.3~kpc, and not longer than 2.5~kpc. Throughout this paper we thus assume $d=1.7_{-0.4}^{+0.8}$~kpc as the most likely value for the distance but will also use 1.3~kpc and 2.5~kpc to derive lower and upper bounds for certain parameters.


The Gaia DR3 catalogue also offers a way to estimate the interstellar extinction by means of the $A_0$ parameter. It is  computed for the monochromatic wavelength 541.4~nm by means of the blue and the red spectra obtained by the Gaia spacecraft. As such, $A_0$ is close to but not exactly the same as the extinction $A_V$ in the Johnson $V$-band \citep{Creevey2022}. Based on the sample of stars used for the histogram construction (Fig.~\ref{fig_distances}), and imposing an extra constrain on the signal-to-noise ratio of the parallax $\geq 7$, we plot the $A_0$ vs. parallax dependency in Figure~\ref{fig_ext_parallax}. Although the latest constraint severely affects the population of the more distant objects, the sample is sufficiently numerous to observe a clear trend. $A_0$ as high as $7-9$~mag are obtained for stars near 2~kpc. If confirmed by more accurate methods of the $A_V$ determination (see in Sections~\ref{sec:Av_spectr} and \ref{sec:sed_mod}), such values would be in line with the very red colour index of Gaia21bty.

\section{Data analysis}
\label{sec:analysis}

\subsection{Variability in the pre-outburst stage}
\label{sec:var_pre}

Gaia21bty showed a few significant dips during the pre-outburst stage, as best evident in the most uniform and numerous data set gathered in the $K_S$ filter by the VVV/VVVX survey. In general, the variability appears to show higher amplitudes at shorter wavelengths. We infer this mainly from the near-infrared data, as shown in Figure~\ref{fig:dips}. This figure shows the $K_S$ and $I$-band light curves dominated by well-defined 0.4-0.5 and 1.5-2~mag dips, respectively. Their minima (indicated by marks) occurred at $JD-2450000\approx$ 6505, 6851 ($K_S$), 7848 ($I$) and 8250 ($K_S$) -- the exact moments are impossible to estimate due to nonuniform sampling and asymmetric properties of the dips. The total duration time of every dip is about 100~days. Assuming that all the four dips were produced by the same obscuring body, e.g. a single disc warp or a dusty cloud, a simple consideration suggests that perhaps for some not yet understood reason the dips occur in 350-400~day pairs roughly every 1340-1400~days.  

Although the $K_S$ and $I$-band observations were obtained during different epochs, similar depth of all three dips observed in the $K_S$-band suggests that it was also the same during the OGLE run. If it is the case, then the stronger light lost in the $I$-band could suggests occultation of a several thousand Kelvin hot source, most likely the central star and/or the very inner (within 0.1~au) disc. This suggests large (60-80~deg) inclination $i$ of the disc. Typical inclination found for AA~Tau-types stars by \citet{McGinnis2015} is about 70~deg and we will assume this value for the disc inclination of Gaia21bty throughout this paper.

\subsection{Light variations during and after the outburst}
\label{sec:var_outb}

Based on Figure~\ref{fig:lc}a, the sudden brightness increase from the quiescent phase ($G\approx20.1$~mag in the second half of 2020) to the maximum ($G=17.28$~mag) lasted no longer than 4 months. 
Assuming the typical quiescent (but also 'uneclipsed') state at $G=20.2$~mag, the total brightening in the $G$-band would be $\Delta G=2.9$~mag. Accordingly, by comparison of the VVV/VVVX data obtained out of dips during 2018 August, and the NTT/SOFI data taken on 2021 May 6, we estimate brightening of $\Delta J=2.14$, $\Delta H=2.06$ and $\Delta K_S=2.24$~mag. Our $\Delta K_S$ is similar as observed in spectroscopically confirmed eruptive stars by \citet{Contreras-Pena2017b}. Comparison of the NEOWISE data gathered in fall 2018 and in spring 2021 results in $\Delta W1 = 1.72$ and $\Delta W2=1.61$~mag. 

In contrast to Gaia17bpi \citep{Hillenbrand2018_ApJ869146H} and Gaia18dvy \citep{SzegediElek2020_ApJ899130S}, it is not clear whether the gradual brightness rise in $W1$ and $W2$ bands starting (perhaps) in the second half of 2019 or at the beginning of 2020 (i.e., about 0.75 year earlier than in the optical Gaia light curve) is significant, or is a typical stochastic variation within the quiescent state. If it is really caused by the {\it outside-in} mechanism considered by the authors for the two aforementioned FUors, then assuming $0.2-1$~M$_{\odot}$ for the stellar mass of Gaia21bty, and that the outburst propagated inward in the disc dynamical time scale (i.e. comparable to the Keplerian period at the disturbed disc annuli), the time difference of 0.75 year suggests that the outburst could have started at $0.5-0.8$~au.

After the maximum around 2021 February -- April, Gaia21bty started to fade. Interestingly, the fading process is not occurring monotonically: based on the Danish telescope and Gaia spacecraft data, in April 2022, i.e., about a year after the maximum, when the star almost reached the pre-outburst level ($G=19.93$~mag), we observed brightening 
that reached $\Delta V = 1.3$~mag. A second brightening was observed in October: the Gaia spacecraft data obtained on 2022 October 3 showed the star at $G=20.37$~mag (typical for the pre-outburst stage), while those obtained barely 18 days later (2022 October 22) showed the star brighter by 1.28~mag. The $V=21.78$~mag measured by the Danish telescope one day earlier is even higher than during the April 2022 brightening. The 2023 data, obtained in February -- April, show continuation of this variability pattern.

Different behaviour is observed by NEOWISE. After a little brightness drop in $W1$ and $W2$ bands in 2021 August 24, which occurred a few months after the maximum, in 2022 the brightness of Gaia21bty increased a bit and roughly stabilised. We emphasize that the first NEOWISE epoch (March 13) was collected simultaneously with our Danish telescope observations, which clearly shows that the optical brightness reached a minimum in 2022 ($V=23.48$~mag). This means that the post-outburst infrared emission is not as strongly correlated with the optical one, as during the first outburst stage. Specifically, the infrared emission has remained strong, while the optical flux has dropped but varied.

In spite of the poor photometric sampling, it appears that the optical $\sim$1~mag brightenings superimposed on the gradual fading are occurring with about a few weeks frequency. They may look like short accretion bursts at first glance, what would mean that the inner disc has not yet returned to the equilibrium. However, a competing explanation by extinction variations will be discussed in Section~\ref{sec:c-c}.

\begin{figure}
    \centering
    \includegraphics[width=1\columnwidth]{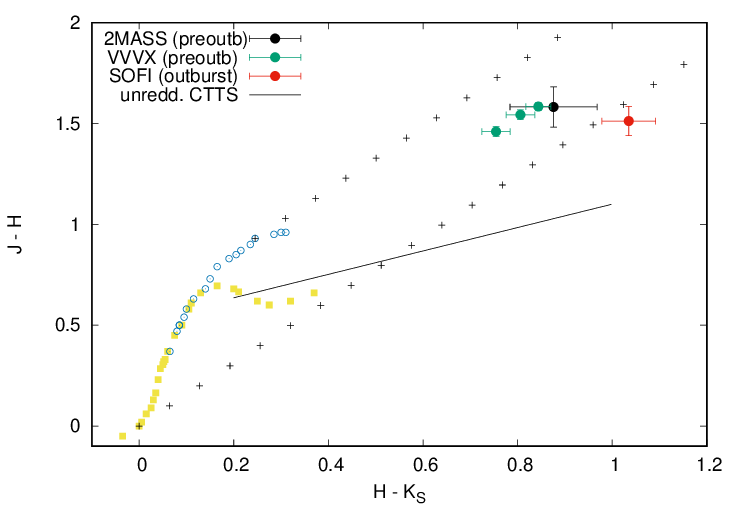}
    \caption{$J-H$ vs. $H-K_S$ colour-colour diagram prepared from pre-outburst (2MASS and VVVX) \& outburst (SOFI) data of Gaia21bty. The squares represent the zero age main sequence stars, while the open circles the giant branch \citep{Bessel1988}. The black line is the locus of unreddened CTTS \citep{Meyer1997}. The two parallel lines formed from pluses represent the reddening path; the step represents the reddening by additional 1~mag in $A_V$.}
    \label{fig_ccd}
\end{figure}

\begin{figure*}
    \centering
    \includegraphics[width=1\columnwidth]{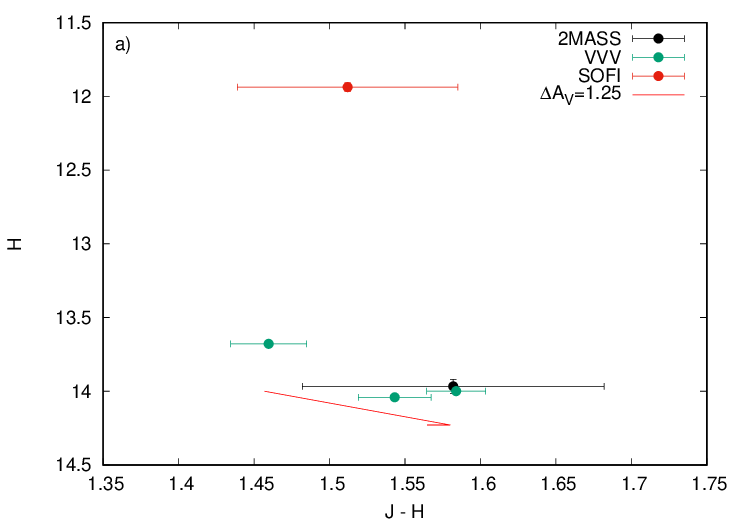}
    \includegraphics[width=1\columnwidth]{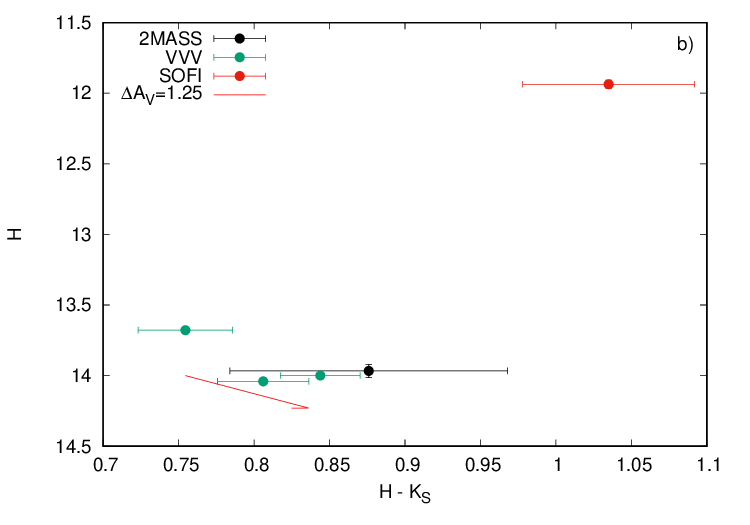}
    \includegraphics[width=1\columnwidth]{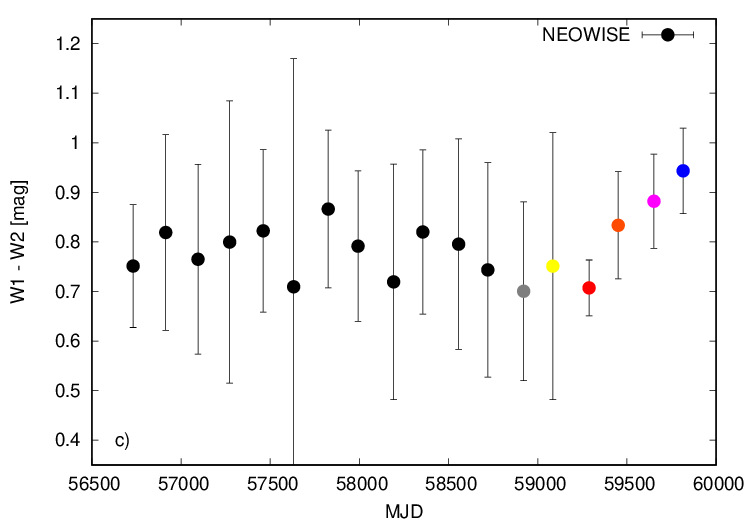}
    \includegraphics[width=1\columnwidth]{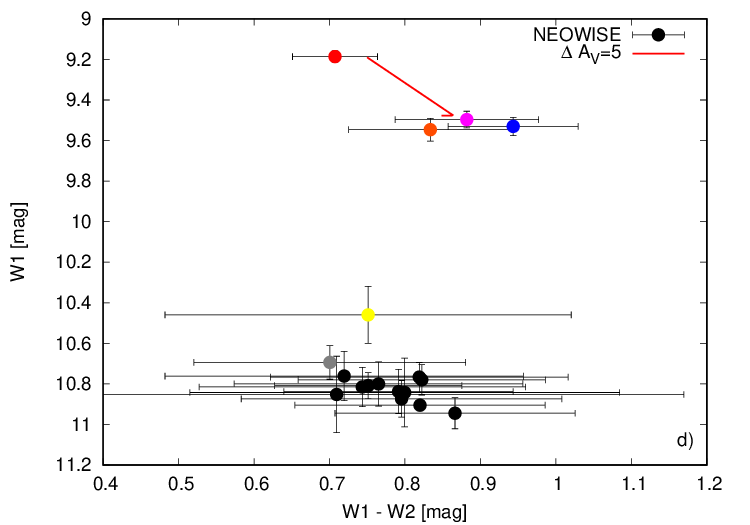}
    \includegraphics[width=1\columnwidth]{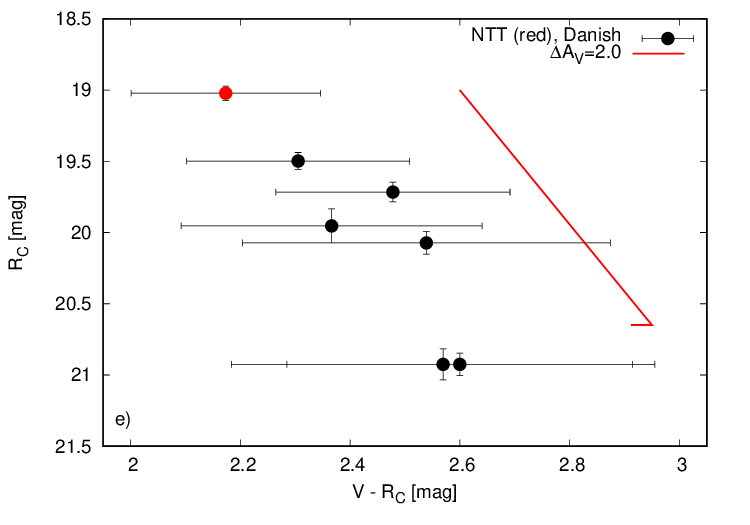}
    \includegraphics[width=1\columnwidth]{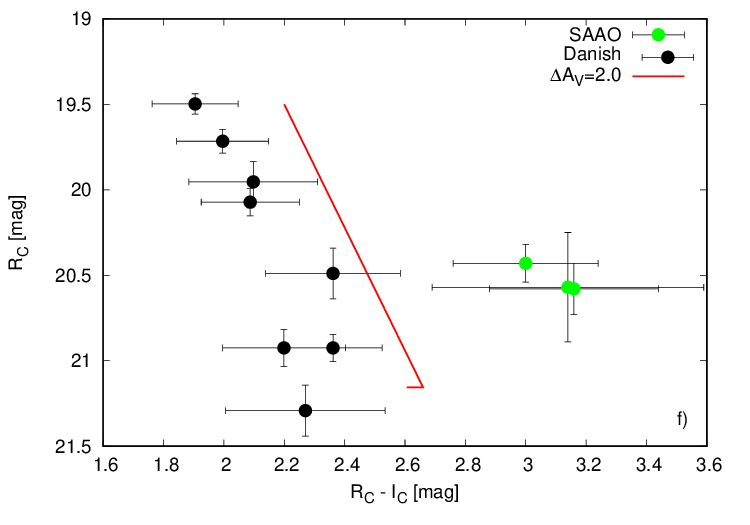}
    \caption{Colour-magnitude diagrams from $JHK_S$ pre-outburst (2MASS, VVV, VVVX) and outburst (SOFI) data of Gaia21bty (panels a and b). Panel c shows $W1-W2$ colour index in function of time, while panel d shows colour-magnitude diagram for pre and outburst phases based on NEOWISE $W1$ and $W2$ data. The outburst points in panels c and d are coloured in the same way to facilitate navigation. The last two panels (e,f) show the $V-R_C$ and $R_C-I_C$ vs. $R_C$ diagrams, constructed from the data obtained in three observatories in the maximum and during the oscillations occurring during the gradual brightness decrease. Red half-arrows indicate reddening vectors for $R_V=3.1$ and given $A_V$ increase, as  indicated in the legends.}
    \label{fig_cmdJHK}
\end{figure*}

\subsection{Colour-magnitude and colour-colour diagrams. Determination of $A_V$ during the quiescent phase}
\label{sec:c-c}

Analysis of the colour variations of Gaia21bty before and during the outburst is not possible in optical bands because no simultaneous multi-filter data are available. Therefore such analysis for the {\it pre-outburst-to-maximum} stage is only possible for the $JHK_S$ and (NEO)WISE data. Accordingly, the $VR_CI_C$ and NEOWISE data enable similar analysis for the {\it maximum-to-quiescence} stage.
\vskip 3pt
{\it Pre-outburst-to-maximum:} In Figure~\ref{fig_ccd} we present the $J-H$ vs. $H-K_S$ colour-colour diagram and the empirical position of unreddened CTTS established by \citet{Meyer1997}. Both the pre-outburst and the outburst colour indices of Gaia21bty indicate strong reddening, as previously inferred from the optical photometry. The analysis made in the previous section indicates that it could be Class~II rather than Class~I (or flat spectrum) object, therefore this diagram can be used to estimate the interstellar visual extinction $A_V$ during quiescence. Assuming $R_V=3.1$ for \citet{Cardelli1989} extinction law, we obtained $A_V=8.0\pm0.5$~mag during the brightest epoch covered with infrared pre-outburst observations (Fig.~\ref{fig_ccd}). 
Values larger by $1.2-1.3$~mag (up to $A_V=9.3$~mag) are derived for epochs representing fainter stages, i.e. the local dip events, as obtained from the 2MASS and the reddest VVV/VVVX data points. Note that the arrangement of the VVV/VVVX colour indices exactly along the extinction vector is in line with the dipper classification (Sec.~\ref{sec:var_pre}). 
We will use other methods for $A_V$ determination in the outburst stage in Sections~\ref{sec:Av_spectr} and \ref{sec:sed_mod}.

Near-infrared colour-magnitude diagrams, infrared colour index vs. time, and colour-magnitude diagram based on NEOWISE photometry are shown in Figures~\ref{fig_cmdJHK}a, \ref{fig_cmdJHK}b, \ref{fig_cmdJHK}c and \ref{fig_cmdJHK}d, respectively. According to the colour-magnitude diagrams composed of the ground-based $JHK_S$ data obtained simultaneously during five nights (Fig.~\ref{fig_cmdJHK}ab), the variability visible during the pre-outburst stage can also be explained by the $A_V$ increase by 1.2-1.3~mag (indicated by the half-arrow in the figure). We note that during the outburst, the NTT/SOFI marked by red, and the NEOWISE data points marked by grey-through-yellow (to indicate the point from 2020 -- the very beginning of the outburst), - to red (to indicate the point from 2022 March 13 -- the maximum itself), do not vary along the extinction path (Fig.~\ref{fig_cmdJHK}abd). Also, the $W1-W2$ colour index appears to be stable within the error bars prior and during the outburst maximum, although not in the post-outburst phase Fig.~\ref{fig_cmdJHK}c). 
\vskip 3pt
{\it Maximum-to-quiescence:} Based on the diagrams composed of the NEOWISE (Fig.~\ref{fig_cmdJHK}d), the NTT/EFOSC2 and the Danish/DFOSC $VR_CI_C$ data (Fig.~\ref{fig_cmdJHK}ef), the brightness evolution from the maximum to the quiescence appears to vary along the extinction path. We note, however, that these optical bands probe similar spectral regions, which makes it difficult to differentiate from changes caused by variable accretion (see e.g. in fig.~7 in \citealt{Szabo2021}).

Interestingly, whereas the Danish/DFOSC data closely follow the extinction vector calculated for $R_V=3.1$, the $R_C-I_C$ colour index observed at the SAAO in August--October 2021, i.e., just after the maximum, is redder by 1~mag at the same $R_C$ brightness (Fig.~\ref{fig_cmdJHK}f). Publicly available technical documentation of these instruments shows that the large deviation cannot be caused by very different filter and/or CCD chip sensitivity curves. Also the SAAO data were reduced in two independent ways and the results are the same within the measurement errors. This apparently real discrepancy could be explained by the assumption that the FUor outburst continues, but the post-maximum light drop and erratic variations are caused by matter composed of different grain sizes, characterised by different {\it total-to-selective} extinction ratios. If it is the case then $R_V=0.6-1$ would be required to explain the changes between these SAAO points and the brightest points obtained by the NTT ($R_C=19.00$~mag) and the Danish telescope ($R_C=19.48$~mag).

This view seems to be supported by 2022 NEOWISE data. They show that the $W1$ and $W2$ magnitudes measured after the maximum (which is indicated by the red point) also faded along the extinction path (i.e. the red, orange, pink and blue points in Fig.~\ref{fig_cmdJHK}d). $\Delta A_V=5$~mag is required to explain the drop observed in the infrared bands. Interestingly, during the lowest brightness noticed in post-maximum NEOWISE data on 2021 August 24, the nearly simultaneous SAAO observations suggested low $R_V$, as discussed above. If these variations are caused by a dusty cloud, its maximal diameter would be of 2~au to produce the observed effect in NEOWISE bands (assuming the maximum inner disc temperature of 6000-7000~K). Furthermore, the first 2022 NEOWISE point was obtained when the $V$-band brightness was also low (23.48~mag). This is an important finding, as we know that this NEOWISE measurement definitely did not coincide with one of the optical brightness peaks. 
All the above suggests variations caused by dust. This would be a scaled-up version of the phenomenon that occurred soon after the HBC~722 outburst and lasted longer than one year \citep{Semkov2017}. This scenario assumes a dust condensation event, caused by powerful disc wind colliding with the outer disc envelope. Such scenario was also proposed for V1057~Cyg to explain periodic dips appearing after the outburst \citep{Szabo2021}. 

One can also speculate that within the framework established in Section~\ref{sec:var_pre}, the post-maximum light drop could directly be linked to the dipper phenomenon observed in the pre-outburst stage. However, the current dip phenomenon would last longer than one year, whereas the dips in the pre-outburst stage lasted no longer that 100 days. 

Finally, as the optical and near-infrared variability amplitude of Gaia21bty is relatively moderate, one can not rule out the possibility that it shares some properties with V2492~Cyg \citep{Covey2011,Aspin2011,Kospal2011,Hillenbrand2013}. According to these authors, after the outburst, V2492~Cyg started to display (probably) quasi-periodic dip and rebrightening events caused by strongly variable, up to $\Delta A_V=35$~mag, extinction. While this caused strong, typically a few magnitude variability in optical and near-infrared bands, the NEOWISE mid-infrared bands are relatively stable (fig.~3 of \citealt{Contreras-Pena2023}). 

\subsection{Determination of $A_V$ during the outburst phase from the infrared spectra}
\label{sec:Av_spectr}

Knowing that the extinction towards FU~Ori is low and relatively well determined ($A_V=1.7\pm0.1$~mag; see in \citealt{Zhu2007,Siwak2018b,Green2019ApJ...887...93G,Lykou2022}), we used another approach to determine the interstellar extinction towards our target. This method relies on a considerable similarity of all bona fide FUor spectra in $JHK$-bands, and follows the idea of \citet{Connelley2018} about checking how much dereddening $\Delta A_V$ should be applied to our near-infrared flux calibrated spectra to match the calibrated spectrum of FU~Ori (but scaled to match the flux level of the target). Both for the NTT/SOFI and IRTF/SpeX spectra we obtained the best match for $\Delta A_V=6.6-6.7$~mag. This results in $A_V=8.35$~mag towards Gaia21bty during the outburst, and the value is consistent with that obtained in Section~\ref{sec:c-c} during the quiescence. 

\begin{figure}
    \centering
    \includegraphics[width=1\columnwidth]{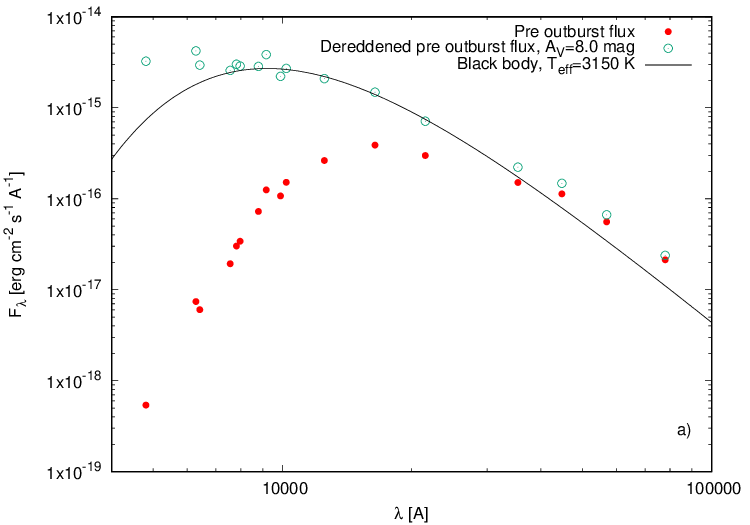}\\
    \includegraphics[width=1\columnwidth]{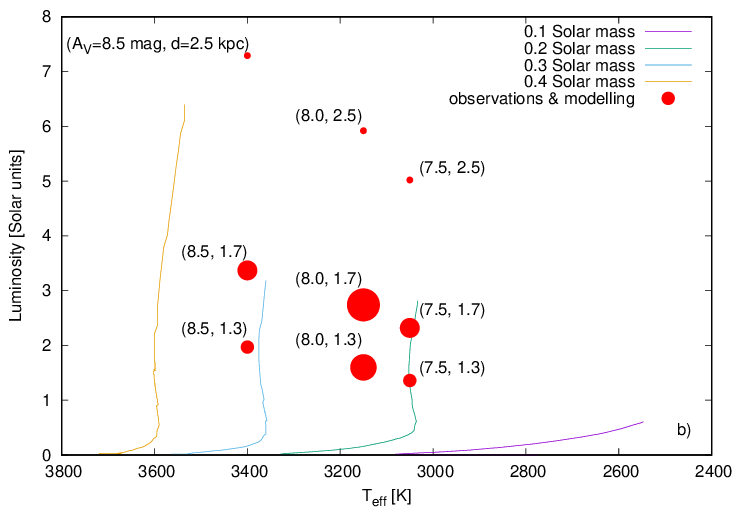}\\
    \includegraphics[width=1\columnwidth]{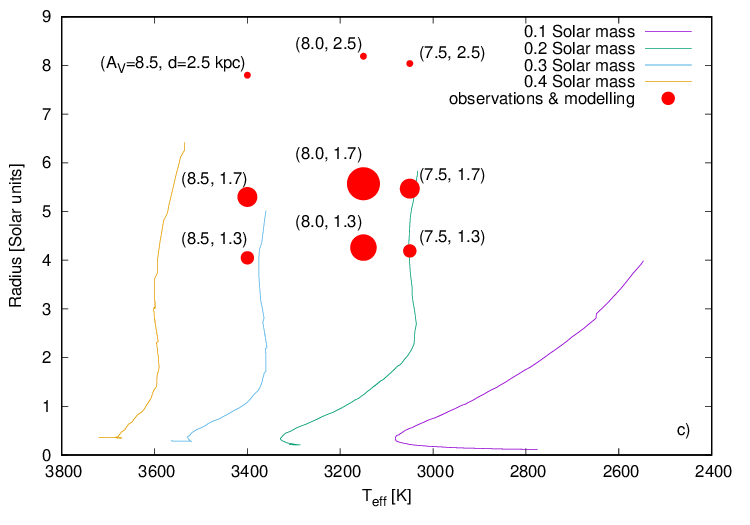}
    \caption{Results of pre-outburst data modelling (a). Position of the star in the H-R diagram based on modelling results and \citep{Pecaut2013} calibration (panels b\&c); those obtained with the assumption of the most optimal set of parameters ($A_V=8.0$~mag, d=1.3 and 1.7 kpc) are indicated by the largest dots.}
    \label{fig_modeling}
\end{figure}

\begin{table}
\centering
\caption{Physical parameters of the progenitor, as obtained in Section~\ref{sec:progenitor}. Those marked by bold represent the most likely set.}
\label{tab:preoutburst_sed}
\begin{tabular}{cccccc}
\hline \hline
$d$~[kpc] & $A_V$~[mag] & $T_{\rm eff}^{\star}$~[K] & $M_{b}^{\star}$~[mag] & $L_{b}^{\star}$~[L$_{\odot}$] & $R^{\star}$~[R$_{\odot}$] \\ \hline
1.3  & 7.5   & 3050     & 4.41 & 1.36  &  4.19       \\
     & 8.0   & 3150     & 4.23 & 1.60  &  4.26       \\ 
     & 8.5   & 3400     & 4.00 & 1.97  &  4.05       \\ \hline
{\bf 1.7}  & 7.5   & 3050  & 3.83  & 2.32    &  5.47 \\
     & {\bf 8.0}   & {\bf 3150} & {\bf 3.65} & {\bf 2.74} &  {\bf 5.57}  \\ 
     & 8.5   & 3400     & 3.42  & 3.37  &  5.30  \\ \hline
2.5  & 7.5   & 3050     & 2.99  & 5.02  &  8.04  \\
     & 8.0   & 3150     & 2.81  & 5.92  &  8.19  \\ 
     & 8.5   & 3400     & 2.58  & 7.29  &  7.80  \\ \hline    
\hline
\end{tabular}
\end{table}

\subsection{Analysis of the quiescent SED. The progenitor}
\label{sec:progenitor}
Using the pre-outburst data and the interstellar extinction determined in Sec~\ref{sec:c-c}, we made an attempt to derive the main physical parameters of the progenitor. We assume full visibility of the star, which may in fact be not completely true given $i=70$~deg.

In the first step, for each band we chose from the full quiescent SED (Fig.~\ref{fig_sed}c) only the points that were not obtained during the dips. Then, we dereddened these selected data by means of the \citet{Cardelli1989} reddening law assuming $A_V=8.0\pm0.5$~mag and $R_V=3.1$ (Sec.~\ref{sec:c-c}). Subsequently, we fitted a Planck function to the near-infrared ($780-2200$~nm) data points. We discarded the shorter wavelengths because they are certainly affected by the hot spots created during the magnetospheric accretion, while the longer by the disc radiation  (Fig.~\ref{fig_modeling}a). As the result we obtained effective temperature of the star of $T_{\rm eff}^{\star}=3150^{+250}_{-100}$~K. The error bars quoted in $T_{\rm eff}^{\star}$ result from the uncertainty in $A_V$: $T_{\rm eff}^{\star}=3050$~K corresponds to $A_V=7.5$~mag, and  $T_{\rm eff}^{\star}=3400$~K to $A_V=8.5$~mag. 
By integrating the respective Planck functions, we calculated three possible values of the bolometric stellar luminosity $L_{b}^{\star}$ at $1.98^{+0.37}_{-0.36}$~L$_{\odot}$ for 1.3~kpc, $3.38^{+0.63}_{-0.61}$~L$_{\odot}$ for 1.7~kpc, and $7.31^{+1.37}_{-1.32}$~L$_{\odot}$ for 2.5~kpc. 
During the Planck fitting to the data dereddened by $A_V=8.0$~mag, we also established preliminary value of the stellar radius $R^{\star}$ at about 4.5, 6 and 8.5~R$_{\odot}$ for $d=1.3$, 1.7 and 2.5~kpc, respectively. 

We also estimated the stellar luminosity and the radius following the procedure described in full in \citet{Fiorellino2021} and applied in Sec.~4.3.2 of \citet{Fernando2022}. Assuming that Gaia21bty shares properties common with Class~II objects, we made use of the \citet{Pecaut2013} {\it spectral type -- effective temperature} 
empirical relation presented in table~6 of their paper. According to this relation, the most likely spectral type is M4, but given the uncertainty in the effective temperature the full possible spread is M2 -- M5. Then, using $m_J=15.138(7)$ corrected for the appropriate value of reddening (i.e. brightened by 2.109~mag for $A_V=$7.5~mag, 2.249~mag for 8.0~mag, and 2.390~mag for 8.5~mag), and bolometric corrections presented by the above authors for respective effective temperatures, we calculated the bolometric magnitude $M_b^{\star}$ of our target for three values of the distance (1.3, 1.7 and 2.5~kpc), and then the respective luminosities ($L_{b}^{\star}$) are $1.60^{+0.38}_{-0.24}$~L$_{\odot}$, $2.74^{+0.63}_{-0.43}$~L$_{\odot}$ and $5.92^{+1.39}_{-0.91}$~L$_{\odot}$. The respective stellar radii calculated with the assumption that the star radiates like a black body are 4.26, 5.57 and 8.19~R$_{\odot}$. These values are similar, although systematically slightly smaller than obtained from the Planck distribution fitting, which may result from the fact that the accretion luminosity, unaccounted by the first method, is now roughly subtracted. We list the results in Table~\ref{tab:preoutburst_sed}. 

Knowing the effective temperature and luminosity, we used the evolutionary tracks for solar composition calculated by \citet{Siess2000} to determine a stellar mass of $0.23^{+0.08}_{-0.03}$~M$_{\odot}$. Based on the same tracks, we determined a formal age of 3600-3700 years only, and (independently of the previous methods) a stellar radius of 4.65-5.75~R$_{\odot}$, which is similar to the one obtained using the \citet{Pecaut2013} calibration. We plot the evolutionary tracks and stellar parameters obtained by means of the \citet{Pecaut2013} relation in Figure~\ref{fig_modeling}b,c. According to the evolutionary tracks, distance higher than 2~kpc seems to be unlikely, as it leads to the large stellar radius that is very atypical for young low-mass stars.

Our results suggest that Gaia21bty could be a very young, rapidly contracting star. Long-term spectrophotometric monitoring in the quiescent state is necessary to check the real influence of the accretion on the blue and longer parts of the spectrum, in order to let us determine the effective temperature and the radius of the star with a better confidence.

\begin{table}
\centering
\caption{Physical parameters of the disc during the outburst maximum. Those marked by bold represent the most likely set. Outer disc radius was set to 0.6~au and inclination to 70~deg during all computations.}
\label{tab:outburst_sed}
\begin{tabular}{ccccc}
\hline \hline
$R_{\rm in}$~[R$_{\odot}$]  & $d$~[kpc] &  $L_{b}^{d}$~[L$_{\odot}$] & $M\dot{M}$~[M$_{\odot}^{2}$~yr$^{-1}$] & $A_V$~[mag] \\ \hline
{\bf2.0}  & 1.3       & 20.2         &  $2.7\times10^{-6}$ & 7.70 \\
          & {\bf 1.7} & {\bf 43.0}   &  ${\bf 5.7\times10^{-6}}$ &{\bf 8.20} \\ 
          & 2.5       & 135.1        &  $1.8\times10^{-5}$ & 8.80 \\ \hline
   4.2    & 1.3       & 10.7         &  $3.1\times10^{-6}$ & 5.40 \\
   5.6    & 1.7       & 17.8         &  $7.2\times10^{-6}$ & 5.35 \\ \hline               
\hline
\end{tabular}
\end{table}

\subsection{Analysis of the outburst SED}
\label{sec:sed_mod}

To study the outburst physics of Gaia21bty in detail, we used the disc model introduced by \citet{Kospal2016} for HBC~722, and then applied to other FUors (for a summary see e.g. \citealt{Szabo2021, Szabo2022}). 
We fitted the outburst SED constructed from photometric data obtained close in time to the brightness peak, in 2021 May (see Fig.~\ref{fig_sed}c). First, the errors of all data points in this SED were conservatively assumed to be 5\%. Then, based on the dipper nature of Gaia21bty, we set the disc inclination to $i=70$~deg. The distance was set to $d=1.7_{-0.4}^{+0.8}$~kpc, while the inner disc radius to $R_{\rm in}=2$~R$_{\odot}$, like for FU~Ori in \citet{Zhu2007}.

Our best fit model resulted in an outer disc radius 0.6~au (note that this is similar to the value estimated by our simple thought experiment in Section~\ref{sec:var_outb}), disc bolometric luminosity $L_{b}^{d}=43.0_{-22.8}^{+92.1}$~L$_{\odot}$, $M\dot{M}= 5.7_{-3.0}^{+12.3}\times10^{-6}$~M$^2_{\odot}$~yr$^{-1}$, and $A_V=8.2_{-0.5}^{+0.6}$~mag. For clarity but especially to show how the results depend on the assumed distance, we list these values also in Table~\ref{tab:outburst_sed}. 
The large uncertainty in $L_{b}^{d}$ and $M\dot{M}$ is caused by uncertainty about the distance. Taking $d=1.7$~kpc, and the stellar mass $0.23^{+0.08}_{-0.03}$~M$_{\odot}$, the mass accretion rate $\dot{M}=2.5^{-0.6}_{+0.4}\times10^{-5}$~M$_{\odot}$~yr$^{-1}$. Based on results listed in Table~\ref{tab:outburst_sed}, larger $\dot{M}$ can be derived for $d=2.5$~kpc, and smaller for 1.3~kpc.

We also explored models with larger inner disc radii, e.g., equal to the stellar radius obtained in Section~\ref{sec:progenitor}. As $d=2.5$~kpc lead to stellar parameters inconsistent with our current knowledge on the early stellar evolution, we considered only those obtained for $d=1.7$~kpc ($R_{\rm in}=5.6$~R$_{\odot}$) and for $d=1.3$~kpc ($R_{\rm in}=4.2$~R$_{\odot}$), in which the stellar radii were inferred from the \citet{Pecaut2013} calibration. Although the fit quality (as measured by $\chi^2$) was now twice better than for the model with $R_{\rm in}=2$~R$_{\odot}$, they resulted in $A_V=5.4$~mag, $M\dot{M}=
3.1\times10^{-6}$~M$^2_{\odot}$~yr$^{-1}$, $L_{b}^{d}$=10.7~L$_{\odot}$, and $A_V=5.35$~mag, $M\dot{M}=7.2\times10^{-6}$~M$^2_{\odot}$~yr$^{-1}$, $L_{b}^{d}$=17.8~L$_{\odot}$, respectively (Tab.~\ref{tab:outburst_sed}). 
These last two $A_V$ values are, however, significantly lower than the $A_V$ values estimated from the outburst spectra (8.35~mag, Sec~\ref{sec:Av_spectr}), or from the colour-colour diagram in the quiescence ($8\pm0.5$~mag, Sec.~\ref{sec:c-c}). Based on these, we accepted $R_{\rm in}=2$~R$_{\odot}$ for the 2021 May outburst SED. Although this is in obvious conflict with the stellar radii determined in the previous section, we can not exclude the scenario, in which due to the enhanced accretion the inner disc matter (a boundary layer) wrapped onto the intermediate latitudes of the star. Other possibility assumes that the disc merges with the bloated star without boundary layer formation, as considered by \citet{Milliner2019}.

We also constructed a SED from data points obtained one year later, in 2022 March 13. In the lack of near-infrared observations, we built this SED from ground-based optical photometry and NEOWISE data. Interestingly, this more recent SED exhibits a shape very different from the one obtained at peak brightness: while in the optical the source dropped significantly since 2021 May, the mid-infrared flux levels remained approximately constant. In order to check whether our simple accretion disc model could reproduce this colour change, we run our best model for $d=1.7$~kpc on the 2022 March SED, fixing the disc geometry parameters. The data points can be well fitted with our disc model, with a luminosity $L_{b}^{d}=38.0$~L$_{\odot}$, $M\dot{M}= 5.1\times10^{-6}$~M$^2_{\odot}$~yr$^{-1}$, and $A_V=10.30$~mag. These results suggest that while the accretion rate slightly decreased after the outburst peak, the main factor behind the detected brightness changes was a sudden rise of the extinction by about 2 mag, which explains also the unexpected colour changes. 

A rise of extinction is a physically plausible explanation during an outburst, since the accretion process can rearrange the structure of the inner disc \citep[e.g.,][]{Mosoni2013}. To explore this possibility, we fitted the 2022 March SED by fixing $A_V$ to 8.20~mag, the value we obtained for the outburst peak epoch in 2021 May, and let $R_{\rm in}$ to vary. We could achieve a fit of similarly quality than before, obtaining the following parameters: $L_{b}^{d}=18.4$~L$_{\odot}$, $M\dot{M}= 5.7\times10^{-6}$~M$^2_{\odot}$~yr$^{-1}$. Based on these results, it is possible to interpret the brightness evolution of Gaia21bty between 2021 May and 2022 March by keeping the accretion rate and the extinction unchanged, but increasing $R_{\rm in}$ from 2.0~R$_{\odot}$ to 4.45~R$_{\odot}$. Physically, this increase in $R_{\rm in}$ might mark a depletion of the innermost part of the disc as the piled-up material falls onto the star during the outburst, converting the inner disc optically thin.  While it is a reasonable alternative scenario in addition to our previous models where $A_V$ changed, it is also possible that we observe the outcome of a simultaneously increasing $A_V$ and a growing $R_{\rm in}$ in the Gaia21bty system.

Our modeling of the SED in 2022 March, either the fixed $R_{\rm in}$ or the fixed $A_V$ scenarios, implies that the extinction remained relatively high, larger or equal to 8.2~mag. We can confirm this conclusion with observations. If the extinction was significantly lower, we probably would be able to detect our target in the $B$-band images taken in the maximum by NTT/EFOSC2 and by Danish/DFOSC during the brightening on 2022 April 12. This is obtained as follows: Gaia21bty is detected with $S/N=7-8$ on each single 300~sec $V$-band images obtained on 2022 April 12, so that a 50~sec exposure would suffice to detect the star with $S/N=3$. According to \citet{Pecaut2013} for Class~II star of M4 spectral type the intrinsic $(B-V)_0=1.53$~mag. The reddening ($R_V=3.1$, $A_V=5.4$~mag) would lead to $E(B-V)=1.76$~mag (vs. 2.61~mag for $A_V=8$~mag), so Gaia21bty would be 3.29~mag fainter in $B$ than in $V$. So in order to reach $S/N=3$ in the $B$-band, we would need to integrate for about 1040~sec. However, the total integration time in the $B$-band was 1500~sec. Thus, the non-detection supports $A_V$ larger than 6.7~mag. 

\begin{figure*}
    \centering
    \includegraphics[width=1.75\columnwidth]{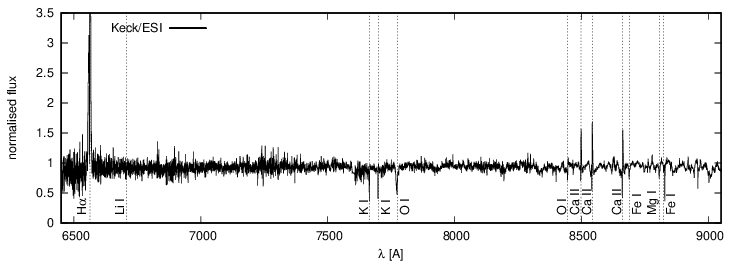}
    \includegraphics[width=1.75\columnwidth]{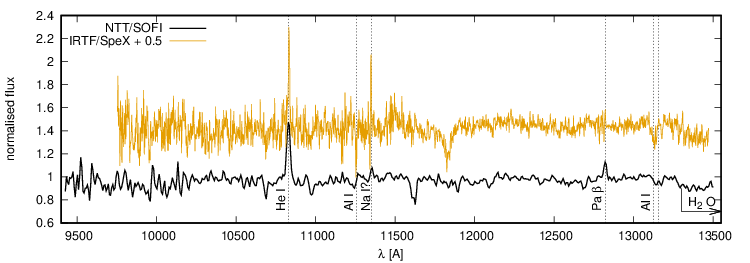}
    \includegraphics[width=1.75\columnwidth]{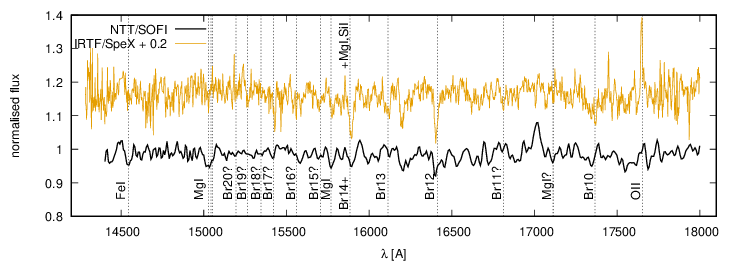}
    \includegraphics[width=1.75\columnwidth]{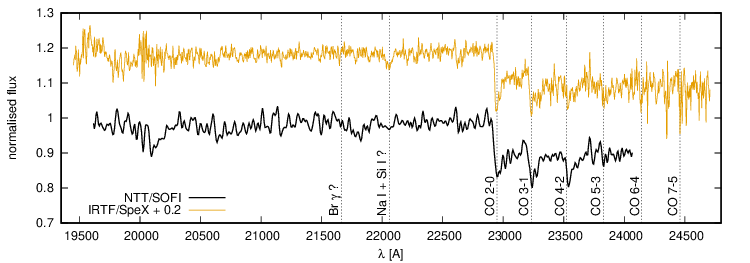}
    \caption{Optical Keck/ESI, and near-infrared NTT/SOFI \& IRTF/SpeX spectra (shifted by 0.5 or 0.2 for clarity), with major detected spectral lines marked with dotted vertical lines. H${\alpha}$ peaks at a normalised flux of 6.5 (so it is cut off here), while the spectral region before this line is dominated by noise.}
    \label{fig_lines}
\end{figure*}

\subsection{Analysis of the outburst spectral properties}
\label{sec:spectral_prop}

In spite of the mass accretion rate typical for FUors, the short maximum duration of the outburst in optical light indicates that Gaia21bty may not be a bonafide FUor. However, our spectra are dominated by features seen typically only in FUors, as described below.

\subsubsection{Comments on individual lines}

In Figure~\ref{fig_lines} we show optical and near-infrared spectra normalised to unity at the continuum level. The major spectral lines we detected are indicated by vertical dashed lines.  

The optical Keck~II/ESI spectrum obtained at the maximum light shows a weak absorption signature of Li~I (6707.76, 6707.97~\AA). It is barely detectable in the original spectrum due to the low S/N, but the visibility can be improved by 3-pixel gaussian smoothing. The manually measured radial velocity (RV) at 2/3 depth of the smoothed line is about $-36\pm2$~km~s$^{-1}$ (Fig.~\ref{fig_vel_lines}a). Sharp and deep K~I doublet (7664.90 and 7698.96~\AA) and the O~I~7774~\AA~ triplet absorptions are also present, and these lines are centered at the same velocity ($-35\pm1$~km~s$^{-1}$) as lithium. This agreement gives hint that this value can be considered as the systemic RV of Gaia21bty with respect to us (see also Sec.~\ref{sec:broadening_function}).

The infrared $J$-band spectra obtained in the maximum by NTT/SOFI, and a second obtained during the brightness decrease by IRTF/SpeX, show metallic absorption lines like Al~I at 1.312 and 1.315~$\mu m$, and water vapor bands in absorption starting with a sharp drop at 1.33~$\mu m$ and leading to the characteristic triangular shape of the $H$-band spectra. The $H$-band spectra appear to be dominated by the Brackett series in absorption -- those visible in one spectrum only, or with uncertain detection in both spectra, are labelled with a question mark in Figure~\ref{fig_lines}. Metals are also present in absorption, such as FeI and Mg~I, one of them (at 1.588~$\mu m$) blending with Br~14-4 and SiI. 
Both $K$-band spectra show the CO bands in strong absorption. Due to the time span of 48 days we compared these spectra to check for difference in CO bands absorption depth, following \citet{Szabo2021}, but we did not notice any significant changes. The IRTF/SpeX spectrum also reveals Na 2.208~$\mu m$ in absorption, however, no firm indications of Ca 2.256~$\mu m$ are present. There is no consensus about the (non)detection of Br$\gamma$ in both spectra. All these spectral features listed so far are common for FUors.

P-Cygni profiles typical for FUors and indicative of outflow from a disc wind, are seen in all three Ca~II IRT lines (8498.02, 8542.09 and 8662.14~\AA; Fig.~\ref{fig_vel_lines}b).  NaD (5889.95 and 5895.924~\AA) seem to exhibit P-Cygni profile as well, but the S/N in this region is too small (about 2) to claim this with full certainty. We also probably detect Na~I line in both J-band spectra, and the line also exhibits P-Cygni profile. 
The K~I doublet and O~I triplet mentioned above have extended blueshifted absorption components to their profiles, though no emission component.

The most prominent spectral lines seen in emission are  H$\alpha$ (6562.82~\AA) and He~I 1.083~$\mu$m, along with weak Pa$\beta$ (12821.6~\AA). 
We also see the infrared O~II line in emission, but only in the spectrum taken in the fading. 

The H$\alpha$ profile is double-peaked with a blueshifted absorption component, most likely due to the superposition of emission and an absorbing wind component (Fig.~\ref{fig_vel_lines}c). The emission components peaks around $-170$ and $+33$~km~s$^{-1}$, and the absorption has a minimum at about $-55$~km~s$^{-1}$. The line peaked $6.5\times$ times over the continuum level which is significantly more than observed ($\sim 1.1-1.4$) in a few FUors, namely HBC~722, V960~Mon, Gaia17bpi and Gaia18dvy during their first outburst phases \citep{Semkov2010,Kospal2016,Miller2011,Hillenbrand2018_ApJ869146H,Park2020,SzegediElek2020_ApJ899130S}. In addition, the H$\alpha$ line in Gaia21bty did not evolve to a pure P-Cygni profile like in the above FUors, the same as it was seen in the spectra taken in the early outburst stages of HBC~722 \citep{Semkov2010, Kospal2016}. 
We speculate that (at least) during the time when our optical spectrum was taken, the accretion rate was still insufficient to drive a disc wind massive enough to develop an even deeper absorption component, capable of producing typical P-Cygni profile, like in the Ca~II triplet (Fig.~\ref{fig_vel_lines}b). This is supported by a comparison of HBC~722 and Gaia17bpi spectra obtained at similar outburst stages, which reveals that the P-Cygni profile in the H$\alpha$ line appeared later in HBC~722 than in Gaia17bpi \citep{Hillenbrand2018_ApJ869146H}. Similar evolution of H$\alpha$ line profile was presented by \citet{Park2020}, who reported that in parallel with the brightness decrease of V960~Mon, the profile evolved from a pure P-Cygni shape to being dominated by an emission component.
We also note that the H$\alpha$ line in Gaia21bty has a strength similar to that observed during the maxima of V899~Mon, which is known to be an ''intermediate'' eruptive star between FUors and EXors \citep{Ninan2015, Park2021}. Continued monitoring should reveal temporal changes occurring on a longer time scale and their possible correlations with the accretion rate in Gaia21bty.

 The resolution of our J-band spectra is insufficient to study the He~I 1.083~$\mu$m line shape in detail, but it appears to be fairly symmetric and shows only the emission component. According to figure~4 in \citet{Erkal2022} who studied the line shape in 117 Class~II sources having disc inclinations determined by ALMA, this profile shape is common for stars observed at high inclinations, which is in line with the dipper nature of Gaia21bty. The authors also mentioned that sources showing only emission component tend to be associated with known jets and outflows, and that they are more common for younger stars. Thus the fact that we still see the He~I 1.083~$\mu$m line in emission, can indicate a powerful outflow, perhaps collimated into a jet.
No more outflow or jet indicators are observed in Gaia21bty, but this is not unexpected given the signal-to-noise ratio of our spectra shortward of 6500~\AA, and the fact that the spectrograph slit was randomly positioned with respect to the putative jet vector. 

We also observed Pa$\beta$ in emission in the NTT/SOFI spectrum obtained at the maximum light, but not in the IRTF/SpeX spectrum obtained 48~days later. 
Taking advantage that the line belongs to the well-known accretion tracers in Class~II stars, we measured the line flux $F_{\rm Pa\beta}$ to independently estimate the mass accretion rate. We decided to perform this check in spite of the fact that Gaia21bty is undergoing enhanced accretion and having properties more typical for FUors.
First, the original flux-calibrated spectra were corrected for the interstellar extinction assuming $A_V=8.35$~mag. Then, following the recipe from section~3.2.5 in \citet{Park2021}, the line flux was measured 100 times with random Gaussian errors multiplied by the observation errors. The standard deviation derived from the all measurements was adopted as the uncertainty of the line flux. 
As the result 
we obtained $F_{\rm Pa\beta}=4.82\pm0.06\times10^{-14}$ erg~s$^{-1}$~cm$^{-2}$. 
Then, following equations number $1-3$ in their paper and using the respective coefficients from \citet{Alcala2017}, for $d=1.7$~kpc, stellar radius of 5.6~$R_{\odot}$ and the stellar mass of 0.23~$M_{\odot}$, we obtained $\dot M=3.1\times10^{-7}$~M$_{\odot}$~yr$^{-1}$, i.e. 2 orders of magnitude less than from the disc modelling. This is giving hint that accretion at the maxiumum brightness of Gaia21bty was no longer magnetospherically controlled, as it would be if Gaia21bty were an ordinary EXor.

\subsubsection{Broadening Function analysis of the Keck/ESI absorption spectrum}
\label{sec:broadening_function}

The better-exposed 8600-10300~\AA~ spectral range shows numerous broad absorption lines, at first sight formed in the disc atmosphere, which could be used to refine the radial velocity of Gaia21bty relative to us. However, the initial radial velocity analysis of a few randomly chosen single lines indicated about three times higher broadening than observed in other classical \citep{Hartmann1985, Kenyon1989} and most recent FUors \citep{Park2020}. 
This is actually not an unexpected effect, as Gaia21bty is an embedded object and its radiation is severely modified by the disc wind. Nevertheless, we decided to amplify the disc component by means of the Broadening Function (BF) method (\citealt{Rucinski2012} and references therein). The method determines the Doppler broadening kernel in the convolution equation transforming a sharp-line spectral template into the observed spectrum using information carried by all absorption lines from the investigated region. This technique was successfully applied for analysis of radial velocities and surface mapping in contact and close binary stars, two CTTS RU~Lup and IM~Lup \citep{Siwak2016}, and also the FUor star V646~Pup. In the latter case \citet{siwak2020} obtained well-defined BFs fairly stable over many years, dominated by the disc profile and showing only a slight signature of the high velocity disc wind outflow in the absorption spectrum. However, the case of Gaia21bty is very different.

\begin{figure*}
    \centering
     \includegraphics[width=1\columnwidth]{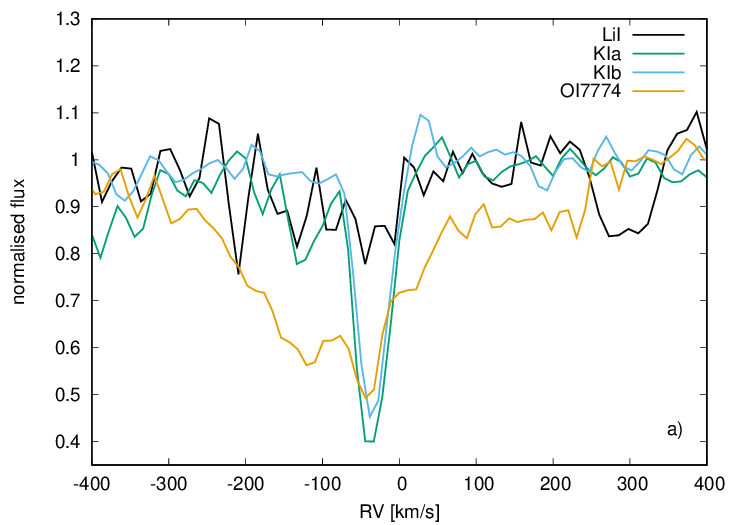}
     \includegraphics[width=1\columnwidth]{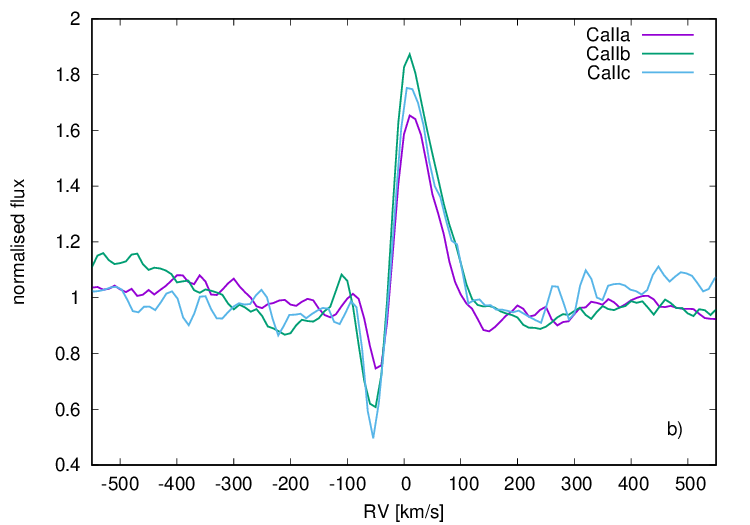}
     \includegraphics[width=1\columnwidth]{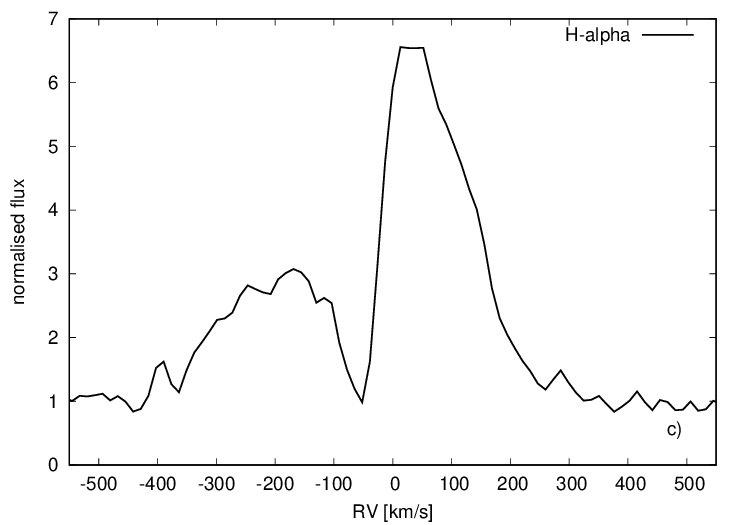}    
     \includegraphics[width=1\columnwidth]{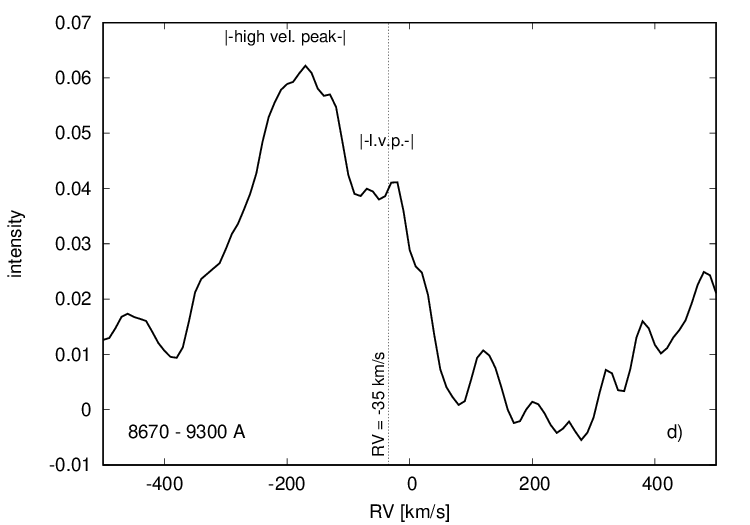}
         \caption{Detailed look onto a few lines observed by Keck/ESI in the maximum. The last panel shows Broadening Functions calculated for the 8670-9300~\AA~spectral region. Preliminary systemic radial velocity, determined by means of the three sharp metallic lines shown in panel (a), is also indicated.}
    \label{fig_vel_lines}
\end{figure*}

We analysed the $8670-9300$~\AA~region using a dozen of synthetic stellar templates. The template spectra were prepared using the PHOENIX database \citep{Husser2013}, they have $T_{\rm eff}=4500-6500$~K with 500~K increment, and $\log g =0.0 - 2.0$ with 0.5 increment. We obtained the best defined BF profiles with the templates having $T_{\rm eff}=5500$~K and $\log g=1-1.5$ (Fig.~\ref{fig_vel_lines}d). Similarly, the analysis of $9556-10256$~\AA~ region resulted in best defined BF profiles obtained with templates having $T_{\rm eff}=4500-5000$~K and $\log g = 0 - 0.5$. These values are in accordance with expectations for FUors, having a super giant stellar spectrum with effective temperature and $\log g$ of consecutive disc annuli decreasing with the increasing distance from the star. The BF profile shown in Figure~\ref{fig_vel_lines}d reveals that the high velocity component (marked as {\it high velocity peak}, peaking at about $-180$~km~s$^{-1}$) dominates over the second peak, marked as {\it l.v.p. (low velocity peak)}. The double-peaked BF shape carries also the information that in contrast to some specific metallic lines (see in Fig.~\ref{fig_vel_lines}a), the great majority of the absorption lines is strongly affected by the disc wind.

The high velocity peak reaching $-300$~km~s$^{-1}$ is formed by the disc wind, and its strength dominates absorption lines similarly like in the embedded FUor V1057~Cyg, as shown in figure~7 by \citet{Herbig2003}, and in figure~10 by \citet{Szabo2021}. We note that the {\it low velocity peak} appears to have similar mean radial velocity as the metallic lines shown in detail in Fig.~\ref{fig_vel_lines}a. This indicates that it represents the pure, unmodified disc absorption. Due to the low brightness of Gaia21bty no high-resolution spectra can be obtained in a close future either to confirm stability of this absorption feature or to study the variation of the line broadening with wavelength, as observed in classical FUors. Unfortunately, the spectral resolution of our $JHK$ spectra is 5-10 $\times$ smaller than provided by ESI, which is insufficient even for rough RV measurements and line profile studies, like above.

\section{Conclusions}\label{sec:summary}


We investigated Gaia21bty, which drew the community's attention on 2021 April 11, as a YSO showing considerable, $\Delta G=2.9$~mag brightening, resembling those observed in other eruptive young stars. The first medium-resolution Keck/ESI visual spectrum obtained at the maximum brightness suggested that it may be a FUor \citep{Hillenbrand2021}. We have been continuing the study by means of public-domain and our own photometric and spectroscopic data. Unexpectedly, Gaia21bty remained in the maximum only for a few months, and around 2021 May-June started to fade in the optical, though has remained bright in the mid-infrared through at least late 2022.

To characterise the progenitor, we constructed the pre-outburst light curve based on images provided by sufficiently deep, public-domain surveys. The historical brightness changed very little, except for a few deep dips. The star appears to be heavily extincted ($A_V=8$~mag), preventing significant detections shortward of the $V$-band. At longer wavelengths, roughly beyond 8-10~$\mu m$, the source characterization is seriously limited or impossible due to the bright nebula surrounding the object. Our classification of the progenitor as a Class~II disc is slightly uncertain, with a flat or even Class~I type spectral energy distribution not ruled out. 

The luminosity of the progenitor is also uncertain, as the distance can be estimated with only a low accuracy based on the distribution of the well-determined parallaxes of other nearby YSOs and YSO-candidates. The most probable value is 1.7~kpc, but the acceptable values lie in the range $1.3 - 2.5$~kpc. 
Planck function fitting to the dereddened data indicates $T_{\rm eff}^{\star}=3150$~K and $L_{b}^{\star}=2.7$~L$_{\odot}$ for the star, which implies a quite large stellar radius of 5.6~R$_{\odot}$. According to the grid of evolutionary tracks by \citet{Siess2000} such a set of physical parameters points to a 0.23~M$_{\odot}$ star that has just entered the Hayashi track. These findings are challenged by our modelling of the outburst SED, as the model obtained with the stellar radius above resulted in $A_V=5.5$~mag, which is inconsistent with that determined during the outburst based on infrared spectra ($A_V=8.35$~mag) and the fact that Gaia21bty is not detected in the $B$-band. According to the disc model, the stellar (assumed to be equal to the inner disc) radius appears to be closer to the canonical value of 2~R$_{\odot}$, which assuming $d=1.7$~kpc resulted in $A_V=8.2$~mag, $L_{b}^{d}=43$~L$_{\odot}$, $M\dot{M}= 5.7\times10^{-6}$~M$^2_{\odot}$~yr$^{-1}$. Assuming the stellar mass of 0.23~M$_{\odot}$, the mass accretion rate would be $\dot{M}\approx2.5\times10^{-5}$~M$_{\odot}$~yr$^{-1}$, which is typical for FUors.

Although the light curve closely resembles those of EXors, the spectral features are typical for FUors. This includes P-Cygni profiles in (at least) the CaII triplet created by a massive disc wind. Although dominated by the emission, similar wind signature can be observed in the H$\alpha$ line too. Strong blueshifted wind signatures dominate the profiles of the metallic absorption lines from the analysed 8670-10300~\AA~ region. The infrared spectra show strong CO absorption band, a series of metallic lines seen in absorption in classical FUors, and the Br$\gamma$ series in absorption. Gaia21bty displays three lines with pure emission profile, namely HeI~1.083$\mu m$, Pa$\beta$ and O~II, and these lines are typically seen in CTTS and EXors, but comparison with other FUor spectra from the literature suggests that this is only because the spectra were obtained in the early stages of the outburst. Note that infrared spectra obtained  48~days later show that Pa$\beta$ emission most likely disappeared, while OII emission appeared or became stronger. 

In conclusion, spectral features typical for classical FUors are definitely prevailing in Gaia21bty.  However, the brightness started to decrease in all bands after barely $4-6$ months of staying in the maximum, which is typical for EXors. As the post-outburst brightness evolution in the colour-magnitude diagrams occurs along the reddening path, and influences mostly the optical wavelengths, we cannot exclude that the outburst continues, but the inner disc and the stellar light is currently blocked. This could be due to dust condensations caused by the disc wind colliding with the outer envelope, as supported by irregular post-outburst photometric behaviour in optical bands, and fairly stable in the infrared. Continued photometric and spectroscopic monitoring should let us understand better the nature of this eruption in the coming years.

\section*{Acknowledgements}
This project has received funding from the European Research Council (ERC) under the European Union's Horizon 2020 research and innovation programme under grant agreement No 716155 (SACCRED). 

We acknowledge support from the ESA PRODEX contract nr. 4000132054.
G. M. and Zs. N. were supported by the J\'anos Bolyai Research Scholarship of the Hungarian Academy of Sciences. G.M. acknowledges support from the European Union’s Horizon 2020 research and innovation programme under grant agreement No. 101004141.

E.F. acknowledges financial support from the project PRIN-INAF 2019 "Spectroscopically Tracing the Disc Dispersal Evolution (STRADE)"

Zs.M.Sz. acknowledges funding from a St Leonards scholarship from the University of St Andrews. For the purpose of open access, the author has applied a Creative Commons Attribution (CC BY) licence to any Author Accepted Manuscript version arising.

This work is based in part on data obtained with the NASA Infrared Telescope Facility, which is operated by the University of Hawaii under a contract with the National Aeronautics and Space Administration.

The data presented herein were obtained at the W. M. Keck Observatory, which is operated as a scientific partnership among the California Institute of Technology, the University of California and the National Aeronautics and Space Administration. The Observatory was made possible by the generous financial support of the W. M. Keck Foundation. The authors wish to recognize and acknowledge the very significant cultural role and reverence that the summit of Maunakea has always had within the indigenous Hawaiian community.  We are most fortunate to have the opportunity to conduct observations from this mountain. 

Based on observations made with ESO Telescopes at the La Silla and Paranal Observatories under programme ID: 105.203T.001. Based on data obtained from the ESO Science Archive Facility.

This paper uses observations made at the South African Astronomical Observatory (SAAO).

This publication was produced within the framework of institutional support for the development of the research organization of Masaryk University.

This publication makes use of data products from the Two Micron All Sky Survey, which is a joint project of the University of Massachusetts and the Infrared Processing and Analysis Center/California Institute of Technology, funded by the National Aeronautics and Space Administration and the National Science Foundation

This work is based in part on observations made with the Spitzer Space Telescope, which was operated by the Jet Propulsion Laboratory, California Institute of Technology under a contract with NASA.

This publication makes use of data products from the Wide-field Infrared Survey Explorer, which is a joint project of the University of California, Los Angeles, and the Jet Propulsion Laboratory/California Institute of Technology, funded by the National Aeronautics and Space Administration.

This research has made use of the NASA/IPAC Infrared Science Archive, which is funded by the National Aeronautics and Space Administration and operated by the California Institute of Technology.

This publication uses data generated via the Zooniverse.org platform, development of which is funded by generous support, including a Global Impact Award from Google, and by a grant from the Alfred P. Sloan Foundation.

We acknowledge ESA Gaia, DPAC and the Photometric Science Alerts Team (\url{http://gsaweb.ast.cam.ac.uk/alerts})

This research has made use of the Spanish Virtual Observatory (\url{http://svo.cab.inta-csic.es}) supported from Ministerio de Ciencia e Innovación through grant PID2020-112949GB-I00.

Special thanks are also due to an anonymous referee for highly useful suggestions and comments on the previous version of the paper.

\section*{Data Availability}

Photometric data extracted from images downloaded from public archives and those obtained in this paper, are presented in the main body of the paper. Extracted spectroscopic data, our individual raw and/or calibrated images, as well as the OGLE data, can be obtained on a request to the corresponding author. All but OGLE photometric data presented in the paper but not presented in the tables were obtained from publicly available catalogues and are therefore not retyped here.



\bibliographystyle{mnras}
\bibliography{gaia21bty} 




%


\bsp	
\label{lastpage}
\end{document}